\global\def\draftcontrol{0}

   \def\versionno{ conics on the mirror quintic }

\catcode`\@=11

\expandafter\ifx\csname draftcontrol\endcsname\relax\global\def\draftcontrol{0} 
\fi 

{\count255=\time\divide\count255 by 60 
\xdef\hourmin{\number\count255} 
\multiply\count255 by-60\advance\count255 by\time 
\xdef\hourmin{\hourmin:\ifnum\count255<10 0\fi\the\count255}} 
\def\draftdate{\number\month/\number\day/\number\year\ \ \ \hourmin } 

\newcommand\makepapertitle{\par

  \begingroup 
    \renewcommand\thefootnote{\@fnsymbol\c@footnote}%
    \def\@makefnmark{\rlap{\@textsuperscript{\normalfont\@thefnmark}}}%
    \long\def\@makefntext##1{\parindent 1em\noindent 
            \hb@xt@1.8em{%
                \hss\@textsuperscript{\normalfont\@thefnmark}}##1}%
     \newpage 
     \global\@topnum\z@   
     \@makepapertitle 
     \thispagestyle{empty}\@thanks 
  \endgroup 
  \setcounter{footnote}{0}%
  \global\let\thanks\relax 
  \global\let\makepapertitle\relax 
  \global\let\@makepapertitle\relax 
  \global\let\@thanks\@empty 
  \global\let\@author\@empty 
  \global\let\@date\@empty 
  \global\let\@title\@empty 
  \global\let\title\relax 
  \global\let\author\relax 
  \global\let\date\relax 
  \global\let\and\relax 
  \def\version{\let\version\@version\@gobble} 
} 
\def\@makepapertitle{%
  \newpage 
   \ifnum\draftcontrol=1 {} 
   \version\versionno 
   \vskip 5.5em%
   \else 
   \hfill\hbox to 3.5cm {\parbox{5cm}{\@pubnum}\hss}%
   \vskip 6.5em%
   \fi 
   \begin{center}%
   \let \footnote \thanks 
      {\hskip -0\textwidth \hbox to 1\textwidth%
        {\centerline{\Large\bf{\noindent%
	\parbox[t]{1.3\textwidth}{\begin{center}\@title\end{center}}}}}}%
     \vskip 1.5em%
     {\normalsize
       \lineskip .5em%
       \begin{tabular}[t]{c}%
         \@author 
       \end{tabular}\par}%
     \vskip 1.5em%
     {\@bstract}%
     \end{center}%
     \vfill
     \@date%
     \vskip 1.5em%
   \par 
} 

\gdef\@pubnum{} 
\def\pubnum#1{%
  \gdef\@pubnum{#1}} 

\gdef\@bstract{} 
\def\Abstract#1{%
  \gdef\@bstract{%
   \parbox{\textwidth-0pc}{%
   \centerline{\bf Abstract}\penalty1000 
   \noindent
   \renewcommand\baselinestretch{1.0} 
   {#1}}} 
} 

\gdef\@email{}
\def\email#1{%
   \gdef\@email{%
   Email: {\tt #1}}
}

\def\ps@paper{\let\@mkboth\@gobbletwo%
     \ifnum\draftcontrol=1 
        \def\@oddfoot{\hbox to \textwidth{\tiny \versionno \hfil\tiny\draftdate}%
        \hskip -\textwidth \hbox to \textwidth{\hfil\rm\thepage\hfil}}%
     \else\def\@oddfoot{\hbox to \textwidth{\hfil\rm\thepage\hfil}} 
     \fi 
     \let\@evenfoot\@oddfoot 
} 

\def\body{\clearpage 
          \pagestyle{paper} 
        } 
\newenvironment{acknowledgments}{%
\vskip 3.25ex 
\addcontentsline{toc}{section}{Acknowledgments}
\noindent {\bf Acknowledgments} 
} 

\def\@version#1{\ifnum\draftcontrol=1 
\typeout{}\typeout{#1}\typeout{} 
\vskip3mm\centerline{\hbox{\fbox{\normalsize{\tt DRAFT -- #1 -- } 
                   {\draftdate}}}}\vskip3mm 
\fi} 
\let\version\@version 
\long\def\eqlabel#1{\ifnum\draftcontrol=1 
                    \tag@false  
                    \tag*{(\theequation) \hbox to -0.2cm{\hspace{0cm}\small{#1}\hss}} 
                    \refstepcounter{equation}  
                    \edef\@currentlabel{\theequation} 
                    \ltx@label{#1}          
                    \else 
                    \label{#1} 
                    \fi 
                    } 
\let\st@bibitem\@bibitem 
\let\st@lbibitem\@lbibitem 
\ifnum\draftcontrol=1 
  \def\@bibitem#1{%
    \st@bibitem{#1}\a@@label{#1}\ignorespaces}
  \def\@lbibitem[#1]#2{%
    \st@lbibitem[#1]{#2}\a@@label{#2}\ignorespaces} 
  \def\a@@label#1{%
    \gdef\a@lab{\smash{\normalfont\small#1}} 
    \ifvmode 
      \if@inlabel 
        \global\setbox\@labels\hbox{%
          \llap{\a@lab\let\a@lab\relax 
                \kern\@totalleftmargin\kern\marginparsep}%
          \box\@labels}%
      \fi 
    \fi} 
\fi 

\documentclass[12pt,letterpaper]{article} 

\usepackage{amsmath,bm,amsfonts,amssymb,array,calc,amsthm,rotating}
\usepackage{epsfig,psfrag} 
\usepackage{graphicx}
\usepackage{color}
\usepackage[colorlinks=true]{hyperref}

\renewcommand\baselinestretch{1.25} 
\renewcommand\section{\@startsection {section}{1}{\z@}%
                                   {-3.5ex \@plus -1ex \@minus -.2ex}%
                                   {2.3ex \@plus.2ex}%
                                   {\normalfont\large\bfseries}} 
\renewcommand\subsection{\@startsection{subsection}{2}{\z@}%
                                   {-3.25ex\@plus -1ex \@minus -.2ex}%
                                   {1.5ex \@plus .2ex}%
                                   {\normalfont\normalsize\bfseries}} 
\renewcommand\subsubsection{\@startsection{subsubsection}{3}{\z@}%
                                   {-3.25ex\@plus -1ex \@minus -.2ex}%
                                   {1.5ex \@plus .2ex}%
                                   {\normalfont\normalsize\it}} 
\renewcommand\paragraph{\@startsection{paragraph}{4}{\z@}%
                                   {-3.25ex\@plus -1ex \@minus -.2ex}%
                                   {1.5ex \@plus .2ex}%
                                   {\normalfont\normalsize\bf}} 
\renewcommand\subparagraph{\@startsection{subparagraph}{5}{\z@}%
                                   {-1.25ex\@plus -1ex \@minus -.2ex}%
                                   {0ex \@plus .2ex}%
                                   {\normalfont\normalsize\it}}

\numberwithin{equation}{section}

\long\def\@makecaption#1#2{%
  \vskip\abovecaptionskip
  \sbox\@tempboxa{{\bf #1:} #2}%
  \ifdim \wd\@tempboxa >\hsize
    {\small\bf #1:} {\small #2}\par
  \else
    \global \@minipagefalse
    \hb@xt@\hsize{\hfil\box\@tempboxa\hfil}%
  \fi
  \vskip\belowcaptionskip}

\renewcommand*\l@section[2]{%
  \ifnum \c@tocdepth >\z@
    \addpenalty\@secpenalty
    \addvspace{.5em \@plus\p@}%
    \setlength\@tempdima{1.5em}%
    \begingroup
      \parindent \z@ \rightskip \@pnumwidth
      \parfillskip -\@pnumwidth
      \leavevmode \bfseries
      \advance\leftskip\@tempdima
      \hskip -\leftskip
      #1\nobreak\hfil \nobreak\hb@xt@\@pnumwidth{\hss #2}\par
    \endgroup
  \fi}
\renewcommand*\l@subsection{\addvspace{.0em \@plus\p@}\@dottedtocline{2}{1.5em}{2.3em}}
\renewcommand*\l@subsubsection{\addvspace{-.2em \@plus\p@}\@dottedtocline{3}{3.8em}{3.2em}}

\def\hepth#1{\href{http://xxx.arxiv.org/abs/hep-th/#1}{{arXiv:hep-th/#1}}}

\def\math#1{\href{http://xxx.arxiv.org/abs/math/#1}{{arXiv:math/#1}}}

\def\arxiv#1#2{\href{http://xxx.arxiv.org/abs/#1}{{arXiv:#1 [#2]}}}

\definecolor{refcol}{rgb}{0.0,0.0,0.2}
\definecolor{eqcol}{rgb}{.2,0,0}
\definecolor{purple}{cmyk}{0,1,0,0}

\gdef\@citecolor{refcol}
\gdef\@linkcolor{eqcol}
\gdef\@urlcolor{refcol}
\def\colorlinkspurple{\gdef\@urlcolor{purple}}
\def\colorlinksblue{\gdef\@urlcolor{blue}}
\def\colorlinksred{\gdef\@urlcolor{red}}

\def\ie{{\it i.e.}} 
\def\eg{{\it e.g.}} 

\def\cf{{\it cf.}}

\def\revise#1       {\raisebox{-0em}{\rule{3pt}{1em}}%
                     \marginpar{\raisebox{.5em}{\vrule width3pt\ 
                     \vrule width0pt height 0pt depth0.5em 
                     \hbox to 0cm{\hspace{0cm}{%
                     \parbox[t]{4em}{\raggedright\footnotesize{#1}}}\hss}}}}

\newcommand\nxt[1]  {\\\fnxt#1} 
\newcommand\bnxt[1]  {\\ $\bullet$ #1}
\newcommand\anxt[1]  {\\ $\ast$ #1}

\def\calc         {{\cal C}} 
 
\def\cale         {{\cal E}}

\def\calh         {{\cal H}} 
\def\cali         {{\cal I}}

\def\call         {{\cal L}} 
\def\calm         {{\cal M}} 
 
\def\calo         {{\cal O}}

\def\calw         {{\cal W}} 

\def\caly         {{\cal Y}}

\def\complex      {{\mathbb C}} 
 
\def\projective   {{\mathbb P}} 
\def\rationals    {{\mathbb Q}} 
\def\reals        {{\mathbb R}} 
\def\zet          {{\mathbb Z}} 
\def\CP{\complex\projective}

\def\del          {\partial} 
 
\def\ee           {{\it e}} 
\def\ii           {{\it i}}

\def\sqr#1#2{{\vcenter{\vbox{\hrule height.#2pt   
 \hbox{\vrule width.#2pt height#1pt \kern#1pt 
 \vrule width.#2pt}\hrule height.#2pt}}}}

\def\Ipp{\mathord{\mathchar "0271 \kern-4.5pt \mathchar"0271}}

\catcode`\@=12 

\begin{document} 


\title{
On the Arithmetic of D-brane Superpotentials.\\
Lines and Conics on the Mirror Quintic}

\pubnum{%
\arxiv{1201.nnnn}{hep-th}\\
NSF-KITP-11-216}
\date{January 2012}

\author{
Johannes Walcher \\[0.2cm]
\it Departments of Physics and Mathematics, McGill University,\\
\it  Montr\'eal, Qu\'ebec, Canada}

\Abstract{
Irrational invariants from D-brane superpotentials are pursued on the mirror quintic, 
systematically according to the degree of a representative curve. Lines are completely
understood: the contribution from isolated lines vanishes. All other lines can 
be deformed holomorphically to the van Geemen lines, whose superpotential is 
determined via the associated inhomogeneous Picard-Fuchs equation. Substantial 
progress is made for conics: the families found by Musta\c{t}\v{a} contain conics
reducible to isolated lines, hence they have a vanishing superpotential. The search
for all conics invariant under a residual $\zet_2$ symmetry reduces to an algebraic problem 
at the limit of our computational capabilities. The main results are of arithmetic 
flavor: the extension of the moduli space by the algebraic cycle splits in the large 
complex structure limit into groups each governed by an algebraic number field. The 
expansion coefficients of the superpotential around large volume remain irrational. 
The integrality of those coefficients is revealed by a new, arithmetic twist of
the di-logarithm: the D-logarithm. There are several options for attempting to explain
how these invariants could arise from the A-model perspective. A successful spacetime
interpretation will require spaces of BPS states to carry number theoretic structures,
such as an action of the Galois group.}

\makepapertitle

\body

\version\versionno

\vskip 1em

\tableofcontents
\newpage

\section{Introduction}

The purpose of this paper is to continue pushing the limit of the calculation of
D-brane superpotentials using the methods developed in \cite{mowa,newissues}. 
The object of study is the value
\begin{equation}
\eqlabel{critical}
\calw(z) = \calw(u;z)|_{\del_u\calw=0}
\end{equation}
of the spacetime superpotential, at the critical point in the open string direction,
compactly denoted by $u$, as a function of the closed string moduli, $z$. Note right
away that \eqref{critical} does not depend on the ambiguous off-shell parameterization 
of the open string moduli space, and is as such a true holomorphic invariant of
the underlying D-brane configuration.\footnote{For a perhaps not over-simplified way
to see arithmetic arising in this context, imagine that $\calw(u)$ is a polynomial 
with integer coefficients, and the superpotential of a supersymmetric theory with 4 
supercharges. Then the supersymmetric vacua and the {\it critical values} 
$\calw|_{\del\calw(u)=0}$,  which are the actual holomorphic invariants encoded in 
$\calw(u)$, generically belong to a finite algebraic extension of the rationals.
The statement of arithmeticy is somewhat different in the context of attractors 
in supergravity \cite{gmoore}, where one looks at critical points of {\it transcendental 
functions} (periods).}

As in \cite{open,newissues}, we have in mind a comparison between three different 
points of view on $\calw(z)$. The mathematically best defined framework is the B-model. 
In that context, $z\in M$ is the complex structure parameter of a family of Calabi-Yau
threefolds $\caly\to M$. We denote the manifold of modulus $z$ by $Y_z$, or simply $Y$ when
$z$ is generic. In standard cases, the D-brane is associated with a family of holomorphic 
vector bundles $\cale$ over $\caly$, or more generally an
object in the derived category $D^b(Y)$ varying appropriately with $z$. The off-shell
superpotential $\calw(u;z)$ which measures the obstructions to deforming $\cale$ in the
open string directions, $u$, as a function of the closed string moduli, $z$, is
given by the holomorphic Chern-Simons functional, or an appropriate extension or 
dimensional reduction thereof for more general objects of $D^b(Y)$. A general effective 
description of the on-shell superpotential involves the (truncated) normal function 
$\nu_\calc(z)$ associated to a family of algebraic cycles $\calc$ that if required can 
be obtained as the algebraic second Chern class of $\cale$. See \cite{mowa} for detailed 
explanations. 

Near a singular point of maximal unipotent monodromy of the family $\caly$, one can obtain
a dual, A-model, point of view on $\calw$. The D-brane there is manufactured 
starting from a Lagrangian submanifold, $L$, of the mirror Calabi-Yau $X$, together 
with a flat connection. While classically the deformations of $L$ are unobstructed, 
worldsheet instanton corrections may induce a superpotential that lifts the D-brane's
moduli space. The critical points of the superpotential
can be identified with unobstructed objects of the Fukaya category. The difficult
problem is to properly count the holomorphic discs with boundary on the Lagrangian 
that give rise to that superpotential.

The third point of view, developed by Ooguri and Vafa \cite{oova}, comes from embedding into
the type IIA/B superstring compactified on $X$/$Y$ for A/B-model respectively. 
One considers the effective two-dimensional theory living on a D-brane partially 
wrapped on $L$ or $\cale$, and extended along a two-dimensional subspace $\reals^2
\subset\reals^4$. According ref.\ \cite{oova}, the superpotential $\calw$ 
not only controls the supersymmetric vacua, but is also related, via its expansion in the 
appropriate limits, to the BPS content of the two-dimensional 
theory. Mirror symmetry relates the choice of $(X,L)$ and $(\caly,\cale)$ and implies
that the superpotentials computed in A- and B-model are identical. Duality with 
M-theory explains the relation to the BPS content of the two-dimensional theory.

The essence of the mirror correspondence is that while the calculations from A-model
or spacetime perspective are forbiddingly difficult in general, the B-model is
 relatively straightforward. On the other hand, the interpretation
in terms of novel geometric invariants is best (though by no means completely) 
understood in the A-model, and most 
interesting, from the spacetime perspective. Interesting mathematics is everywhere.
In this paper, we will present new B-model calculations whose successful A-model 
and space-time interpretation could force a significant extension of the reach of 
these models, into number theory.

Detailed calculations in open string mirror symmetry are now available in a variety
of situations. Motivated and guided by a number of works involving non-compact
manifolds \cite{agva,akv,mayr,lmw}, a quantitative mirror correspondence involving
D-branes on the quintic was established in \cite{open}. Further works involving compact
manifolds include \cite{krwa,knsc,joso,ghkk,ahmm,agbe,japs1,japs2,latest}.
In all these examples, the underlying manifold was selected from the beginning
of the long and well-known list of complete intersections in toric varieties,
for instance hypersurfaces in weighted projective spaces. (The noteworthy
exception is \cite{japs2}, which deals with Pfaffian Calabi-Yau manifolds.)
The choice of cycle in the B-model followed by exploiting divisibility properties 
of weights of specific monomials. In a sense, these D-branes were as close as they 
could be, to the ``toric branes'' that are customarily studied in the context of 
non-compact examples. Somewhat by accident, a subset of those cycles turned out to 
be relevant as the mirror of real slices of the A-model manifold. 

In this work, we return to the quintic Calabi-Yau, with a somewhat different
rationale for selecting the D-branes. As far as A-model is concerned, methods 
for constructing Lagrangian submanifolds of compact Calabi-Yau manifolds are 
in short supply. As far as the B-model, holomorphic vector bundles are  
much easier to produce, perhaps surpassed in simplicity only by matrix
factorizations. The simplest constructions, however, pull vector bundles back
from projective space, which results in rather boring superpotentials (at least
on-shell). Matrix factorizations have the additional disadvantage that they do 
not come with a readily usable version of holomorphic Chern-Simons. Since to
obtain an interesting holomorphic invariant, what we really need is a non-trivial
algebraic cycle class. So we might as well construct the representative cycle
$\calc$, directly, and then calculate as in \cite{mowa}. Although this seems like
a reasonable approach to finding new D-branes, it is not systematically developed. 
So a large initial portion of this work is concerned with identifying 
appropriate $\calc$.

There are several motivations for pursuing this direction. First of all, these
methods will definitely take us further away from the set of toric branes, and
we can prepare ourselves for unexpected new phenomena. (As mentioned above, the
calculations in \cite{japs2} are outside the toric realm. However, the complication
there is introduced in the bulk, \ie, at the level of the Calabi-Yau,
while the D-branes follow the simpler pattern,
conjecturally related to the real A-branes.) In due course, these results
will shape expectations in investigating A-branes and their invariants.

Another broad motivation for this work is a more systematic exploration of the
set of all possible D-branes for fixed closed string background, from the holomorphic
point of view. This is related on the one hand to speculations about background
independence in this context \cite{newa}. On the other hand, a better 
overview over the set of
all D-branes might also be important for realizing open/closed string correspondence
on compact manifolds. In the context of the topological string, the invariant 
holomorphic information contained in the 
superpotential could be the minimal amount necessary.

To organize the advance, we recall again that on general grounds, all possible
on-shell superpotentials in the sense of \eqref{critical} are realized geometrically
as truncated normal functions. On families of
Calabi-Yau threefolds, we are looking for the image of the Chow group
${\rm CH}^2(Y)$ of algebraic cycles modulo rational equivalence, under the 
Abel-Jacobi map. A natural filtration on ${\rm CH}^2(Y)$ is the minimal degree of a 
curve representing a given cycle class. This minimal degree is our organizing
principle. In this paper, we will proceed up to degree 1 and 2 on the mirror
quintic, which as it turns out are already immensely interesting.

We begin in section \ref{vangeemen} with a review of what is known about lines on the
mirror quintic, and then proceed to calculate the inhomogeneous Picard-Fuchs
equation associated with van Geemen lines. The dissection of conics in section 
\ref{conics} is perhaps hard to follow, so we have attempted a shorter
summary in section \ref{sofar}. The first part of section \ref{largecomplex} is 
warmly recommended, as well as a glance at eq.\ \eqref{ffirst}. Section 
\ref{largevolume} contains the main results, and section \ref{discussion} is a 
best attempt at an interpretation.

\section{Lines on the Mirror Quintic}
\label{vangeemen}

Investigations into low degree curves on the quintic have a long history reaching
back before the beginning of mirror symmetry, and provided important information
regarding the latter's enumerative predictions. Instead of the generic quintic, we 
are here concerned with curves on the special one-parameter family of mirror quintics. 
This family is related to the vanishing locus of the polynomial
\begin{equation}
\eqlabel{WW}
W= \frac {x_1^5}5 + \frac{x_2^5}5+\frac{x_3^5}5 + \frac{x_4^5}5
+ \frac{x_5^5}5 - \psi x_1x_2x_3x_4x_5
\end{equation}
in five homogeneous complex coordinates $(x_1,\ldots,x_5)$, and the one parameter,
$\psi$. We denote by $Y$ the generic quintic in $\projective^4$. By $Y_\psi$ we 
denote the member of the Dwork family $\caly\to M$ for fixed $\psi$,
given by $\{W=0\}\subset\projective^4$. The actual mirror quintic is of
course the resolution of the quotient of $Y_\psi$ by $(\zet_5)^3$. This is
useful to keep in mind, but will play only a minor role in the present
discussion.

\subsection{2875 = 375 + 2500}

The space of lines on the one-parameter family of mirror quintics has been investigated
thoroughly by Musta\c t\v a \cite{mustatathesis}, building on the earlier work 
\cite{albanokatz}. The main results of \cite{mustatathesis} is the following: for
fixed generic $\psi$, the quintic $Y_\psi$ contains precisely $375$ isolated lines, and
$2$ (isomorphic) families of lines, each parameterized by a genus $626$ curve. One
of the isolated lines is the coordinate line
\begin{equation}
\eqlabel{coordinate}
x_1+x_2=0 \,,\qquad
x_3+x_4=0\,,\qquad
x_5=0\,,
\end{equation}
while the others are obtained by either permuting the $(x_1,\ldots,x_5)$, or inserting
a fifth root of unity in the first two equations. This leads to the count $5!/2^3\cdot 5^2
=375$.

Special members of the families can easily be written down. If $\omega$ is a non-trivial
third root of unity, and $(a,b)$ satisfy the equations
\begin{equation}
\eqlabel{mostly}
a^5+b^5=27\,,\qquad \psi ab = 6
\end{equation}
then the line
\begin{equation}
\eqlabel{perhaps}
\begin{split}
x_1+\omega x_2+\omega^2 x_3 &=0 \\
a(x_1+x_2+x_3) - 3 x_4 &=0\\
b(x_1+x_2+x_3)-3 x_5 &=0
\end{split}
\end{equation}
lies on the quintic $Y_\psi$. (This is easiest to see by parameterizing solutions of
\eqref{perhaps} as
\begin{equation}
(x_1,x_2,x_3,x_4,x_5) = (u+v,u+\omega v,u+\omega^2 v,au,bu)
\end{equation}
where $(u,v)$ are homogeneous coordinates on $\projective^1$. Then plugging this 
into \eqref{WW}, and using $1+\omega+\omega^2=0$, gives $u^5(3+a^5+b^5-5\psi ab)+
u^2 v^3(30-5\psi ab)=0$
which directly yields \eqref{mostly}.)

Taking into account the phase and permutation symmetries, one obtains a set of
$5000$ lines, called van Geemen lines. This being more that the number of lines
on the generic quintic threefold, which is $2875$, was historically important
because it allowed the conclusion that there exist families of lines on the generic
member $Y_\psi$ of the family \eqref{WW}. The structure of the families at fixed
$\psi$, as mentioned above, was worked out only more recently, and consists of
two curves of genus $626$.

Anticipating results of our Abel-Jacobi calculations in the next subsection, we 
note how it will distinguish the two families of lines: exchanging $a$ and $b$ is
equivalent to exchanging $x_4$ and $x_5$, from which the holomorphic three-form
and hence the normal function, and superpotential, pick up a minus sign. In a
slightly different way, changing the choice of third root of unity, \ie, the
transformation 
\begin{equation}
\eqlabel{firstgalois}
\omega\mapsto\omega^2
\end{equation}
is equivalent to exchanging $x_2$ and $x_3$, and hence also inverts the Abel-Jacobi
image.

The global structure of the families of lines, with varying $\psi$, was also worked
out in \cite{mustatathesis}: the curves parameterizing the families containing the
van Geemen lines fit together to a single smooth irreducible surface, whose Stein
factorization (\ie, the collapse of the connected components in the fibers) 
gives a double cover of $\psi$-space, with branch points at $\psi=0$, 
and $\psi^5=\frac{128}3$.
This is the discriminant locus of the equations \eqref{mostly}. In particular, the
two families that are distinguished for fixed $\psi$ are exchanged as one moves
around in the complex structure moduli space. Quite importantly however, $\psi=\infty$
is not a branch point, so in particular, the choice of third root of unity $\omega$
is a good invariant to distinguish classes of D-branes, in the large complex structure
limit.

Let us record this as the first instance of an intriguing observation: the mapping
\eqref{firstgalois} is nothing but the Galois group of the number field generated
by $\omega$ (which is the imaginary quadratic number field $\rationals(\sqrt{-3})$).
The statement about Abel-Jacobi means that {\it the space-time superpotential 
furnishes a non-trivial representation of the Galois group of the number field over
which the D-brane is defined.} Following our sober discussion, this might not seem
so very surprising. But it has some astonishing consequences that we will explore 
later on.

To close, we repeat here the count of lines which shows that the isolated lines
and the two families account for all rational curves of degree $1$ on the family of
mirror quintics. (That is, for generic values of $\psi$. At $\psi=0$, for instance,
all lines belong to families, as explained in \cite{albanokatz}, and exploited 
frequently.) The families contributing with the Euler characteristic of their
parameter space gives,
\begin{equation}
2\cdot (2\cdot 626-2)+375 = 2500 +375=2875
\end{equation}

\subsection{Inhomogeneous Picard-Fuchs equation}

We need to recall a minimum of material from \cite{mowa,newissues}: if $Y$ is a 
Calabi-Yau threefold, and $C\subset Y$ a holomorphic curve, it makes a contribution
to the superpotential \cite{witten}
\begin{equation}
\eqlabel{understanding}
\calw(z) = \int^C\Omega
\end{equation}
This depends on the complex structure parameter $z$ via the choice of a holomorphic
three-form $\Omega$ on $Y$, which is to be integrated over a three-chain $\Gamma$
ending on $C$. We have written \eqref{understanding} with the understanding that
the actual physical invariant quantities are the tensions of BPS domain walls
(or masses of BPS solitons), which are given by the difference of superpotential
values at the critical points, so $\Gamma$ is then the three-chain interpolating
between two homologous holomorphic curves.

The reason that \eqref{understanding} makes sense even when $C$ is non-trivial
in homology is that we calculate $\calw$ as a solution of the inhomogeneous
Picard-Fuchs equation,
\begin{equation}
\call\calw(z) = f(z)
\end{equation}
where $\call$ is the Picard-Fuchs differential operator of the family $(\caly,\Omega)$.
Since $\call\Omega=d\beta$ is an exact form, the inhomogeneity $f(z)$ originates
from integrating $\int_C\beta$, together with some contribution from differentiating
$C$. Both are clearly local and meaningfully associated to $C$, whether homologically
trivial or not.

With respect to the standard choice of $\Omega$, the Picard-Fuchs operator of the
quintic mirror has the form
\begin{equation}
\eqlabel{standard}
\call=\theta^4-5z(5\theta+1)(5\theta+2)(5\theta+3)(5\theta+4)
\end{equation}
(where $z=(5\psi)^{-5}$ and $\theta=\frac{d}{d\ln z}$.)
The inhomogeneity $f(z)$ was calculated in \cite{mowa} for the Deligne conics
$C_{\pm}$ given by $x_5^2=\pm\sqrt{5\psi} x_1x_3$, within the plane $P=
\{x_1+x_2=0\,, x_3+x_4=0\}$, with the result
\begin{equation}
\eqlabel{deligne}
f_\pm(z) = \pm \frac{15}{32\pi^2} \sqrt{z}
\end{equation}
Since the line \eqref{coordinate} is residual to those conics in the intersection
of $P$ with $Y$, and since, on general grounds, the inhomogeneity associated to 
$P\cap Y$ vanishes, we can conclude immediately that $f(z)=0$ for any of the
$375$ isolated lines.

Another general Hodge theoretic result is that curves that can be holomorphically
deformed into each other give rise to the same normal function. Mathematically, this
is the statement that ``algebraic equivalence implies Abel-Jacobi equivalence''
(a statement valid for curves on Calabi-Yau threefolds). Physically, finite holomorphic
deformations correspond to open string moduli, which are flat directions of the
superpotential.

Given this, we conclude that the two families of lines containing the van Geemen
lines map under Abel-Jacobi each to a single point in the intermediate Jacobian.
It is therefore sufficient to calculate just for the van Geemen lines. Also, as
anticipated above, the images of the two families differ just by a sign.

To calculate $f(z)$ for the van Geemen lines, we may proceed as in \cite{mowa,newissues}.
The key feature to exploit is that any of the lines is part of the intersection of $Y$ 
with a plane, and that the calculation of $\int_C\beta$ localizes to the intersection
points of $C$ with the residual quartic in that plane. The difference to 
\cite{mowa,newissues} is that here there is actually a two-parameter family of planes 
containing any given $C$, so we can make any choice that seems convenient.

The details are straightforwardly executed, and we obtain the inhomogeneity 
associated with the van Geemen lines,
\begin{equation}
\eqlabel{vginh}
f_{\rm van\; Geemen}(z) = \frac{1+2\omega}{4\pi^2}\,\cdot\,
\frac{32}{45}\,\cdot\,
\frac{\frac{63}{\psi^{5}}+\frac{1824}{\psi^{10}}-\frac{512}{\psi^{15}}}
{\Bigl(1-\frac{128}{3\psi^5}\Bigr)^{5/2}}
\end{equation}
($z=(5\psi)^{-5}$). Notice that as $\omega$ is a non-trivial third root of unity,
the inhomogeneity has an overall factor $\sqrt{-3}$ multiplying a function with
a power series expansion around $\psi=\infty$ with rational coefficients. The
main theme of this paper is to investigate irrationalities in the expansion
of the solutions of inhomogeneous Picard-Fuchs equations. As an overall factor, 
the irrationality might seem rather mild in the present case. This is however 
dictated by the anticipated sign change under the Galois action $\sqrt{-3}
\mapsto-\sqrt{-3}$. Later examples will be more complicated, and also the
solutions of the inhomogeneous Picard-Fuchs equation associated to \eqref{vginh}
will already be quite illuminating, see section \ref{largevolume}.

\section{Conics on the Mirror Quintic}
\label{conics}

The basic framework to search for conics on the quintic is easy to describe, 
following Katz \cite{katz}: the moduli space of conics in $\projective^4$ is fibered 
over the Grassmannian $G(3,5)$ of projective planes in $\projective^4$. The fiber
over a plane $A\cong\projective^2\subset\projective^4$ is given by the conics
in a fixed $\projective^2$, spanned by the monomials of degree $2$ in three 
homogeneous coordinates on $\projective^2$, and isomorphic to a copy of 
$\projective^5$. We denote a conic in a fixed $\projective^2$ by $B$. The
conic $B\subset A\subset \projective^4$ is contained in the quintic $Y$ precisely
if
\begin{equation}
\eqlabel{practice}
Y\cap A = B\cup C
\end{equation}
where $C$ is a cubic curve in $A$. 

\subsection{Overview}

In practice, the equation \eqref{practice} means the following: the plane $A$ is 
defined as the vanishing locus of two linearly independent linear equations in the 
five homogeneous coordinates, $x_1,\ldots,x_5$ of $\projective^4$. Up to taking linear 
combinations of those two equations, there are $6$ independent parameters entering these 
equations, which are just (local) coordinates on $G(3,5)$. The equations for $A$ being 
linear, and non-degenerate, they can be solved for two of the five $x_i$'s, say $x_4$ 
and $x_5$. The result can be substituted in the quintic polynomial defining $Y$, 
yielding a quintic polynomial, $p_5$, in $3$ variables. Note that there are $21$ different 
quintic monomials in $3$ variables. 

The conic $B\subset A$ is given as the vanishing locus of a quadratic polynomial $p_2$
in three variables, say $(x_1,x_2,x_3)$, and depends on $6$ homogeneous parameters. Likewise,
the residual cubic is given by a cubic polynomial $p_3$, and depends on $10$ parameters.
The equation \eqref{practice} then is the vanishing of the coefficients of the $21$ 
independent quintic monomials in
\begin{equation}
\eqlabel{linearly}
p_5 - p_2\cdot p_3
\end{equation}
Since we may fix the scale of either $p_2$ or $p_3$ arbitrarily, there are $6+6+10-1=21$ 
independent parameters entering those $21$ equations. (Note that the $10$ parameters for 
$C$ enter linearly in \eqref{linearly}, which may therefore a priori be reduced to a 
system of $11$ equations in $11$ variables. This is the more standard dimensionality 
of the counting problem.) Generically then, we expect a finite number of solutions. 
This is in fact true, and there are $609250$ conics on the generic quintic \cite{katz}.

For special quintics, for example a member $Y_\psi$ of the one-parameter family of quintics 
\eqref{WW}, there will be some number of isolated solutions, and some number of 
continuous families. There can also be conics that are reducible to two intersecting lines. 
When counted appropriately, all these will add up to $609250$. In 
this paper, our main focus is not on counting solutions, but on performing calculations 
for particular conics that deform with $Y_\psi$ as $\psi$ is varied. Conics that
are isolated for fixed $\psi$ will deform to one-parameter families, while families
that exist at fixed $\psi$ can either deform as families or be lifted to isolated
solutions. Globally these local branches of solutions will fit together to various
components of the ``relative Hilbert scheme'' of conics $\calh_{\rm conics}\to M$ on the 
one-parameter family of quintics \eqref{WW}, $\caly\to M$. 

The goal in this section is to identify an interesting subset of components of $\calh_{\rm 
conics}$. In the next section, we will study the branch structure around the large complex 
structure limit, $\psi\to\infty$. To simplify our life, we will neglect obstructed families 
of conics, avoid the singular loci, and all other phenomena that occur at special values of 
$\psi$. 

A fair number of solutions of \eqref{practice} can be found by exploiting the symmetries of
the problem. The full symmetry group $G$ of \eqref{WW} consists of the phase symmetries 
multiplying the $x_i$'s and $\psi$ by fifth roots of unity, and the symmetric group $S^5$ 
that acts by permuting the $x_i$'s,
\begin{equation}
\eqlabel{symmetries}
(\zet_5)^4 \to G \to S^5
\end{equation}
The two subgroups play a somewhat different role in the problem. To construct the mirror 
quintic, we are ultimately interested in dividing out by the subgroup $(\zet_5)^3
\subset(\zet_5)^4$ fixing $\psi$. This means that we should be looking at orbits of curves 
under the group $(\zet_5)^3$, and a non-trivial stabilizer contributes an additional 
factor at the very end of the calculation. On the other hand, no subgroup of $S^5$ will be 
gauged, and a curve with non-trivial stabilizer in $S^5$ is not special in any other
way.

At a more practical level, the phase symmetries act diagonally on the variables 
parameterizing $A$, $B$, and $C$, and dividing out by them does not reduce the dimensionality
of the problem, but merely the degree (which is quite helpful anyways, of course!).
The permutation symmetries act non-diagonally, and can reduce both the dimensionality
and the degree. It is a good idea to keep track whether the subgroup of interest acts
with unit determinant on the $x_i$'s or not. If it does, one might expect the solutions
of the reduced problem to still be isolated, although this is neither
necessary nor sufficient in general. Also, we may point out that a conic that
is isolated as a solution invariant under a particular symmetry could in fact sit in a 
family of conics the generic member of which breaks that symmetry. 

We will return to pointing out these, and many more, features of the space of
conics after we have presented a few explicit solutions.

\subsection{\texorpdfstring{$S^3$}{S3}-invariant conics}

To begin with, one may look for conics that are invariant under permutation of 
three of the five homogeneous coordinates of $\CP^4$, which we choose to be 
$x_1,x_2,x_3$, see \cite{mustata}.
(If our concern were counting conics, we would of course have to account for that choice.)
We parameterize the equations for the plane as follows
\begin{equation}
\eqlabel{plane1}
A : \left\{ \begin{array}{c}
a_1(x_1+x_2 + x_3) + x_4\, \\
a_2(x_1+x_2+x_3)+x_5 
\end{array}\right\}
\end{equation}
and solve for $x_4$ and $x_5$. Note that this means in principle that we are working in a 
specific open patch of the full moduli space. One can check that the solutions in the
other patches precisely serve to compactify the families that we shall write down below.
The conic $B\subset A$ is given by
\begin{equation}
\eqlabel{conic1}
B : \{ x_1^2 +x_2^2 +x_3^2 + b_1(x_1x_2+x_1x_3+x_2x_3) \}
\end{equation}
where we have gauged the coefficient of $x_1^2+x_2^2+x_3^2$ to $1$. This is again only an open
patch, but it captures all solutions. For once, we display the residual cubic:
\begin{equation}
\eqlabel{cubic1}
C : \{c_1\,(x_1^3+x_2^3+x_3^3) + c_2\,(x_1^2x_2+ x_1^2 x_3+x_2^2x_1+x_2^2x_3+x_3^2x_1+x_3^2x_2)
+ c_3\, x_1x_2x_3 \}
\end{equation}
which depends on three parameters. Finally, there are five symmetric polynomials of degree 5 in 
three variables, giving rise to as many equations for the six variables $a_1,a_2,b_1,c_1,c_2,
c_3$. Thus we see that we generically expect a one-parameter family of solutions (for fixed
$\psi$). Writing out those equations explicitly, we see that four of them can be solved linearly
for $c_1, c_2, c_3$, and $b_1$ in terms of $a_1$ and $a_2$. For example,
\begin{equation}
b_1 = \frac{1-2\psi a_1a_2}{1-\psi a_1a_2}
\end{equation}
The remaining equation is
\begin{equation}
\eqlabel{S3invariant}
1-  a_1^5 - a_2^5 + 5 \psi^2 a_1^2 a_2^2 - 5 \psi a_1a_2
\end{equation}
Thus, for fixed $\psi$, picking any solution of \eqref{S3invariant}, the intersection of
the quintic $Y_\psi$ with the plane \eqref{plane1} decomposes as the union of
the conic \eqref{conic1} and the cubic \eqref{cubic1}. This is the solution found by
Musta\c{t}\v{a} \cite{mustata}. 

For completeness, and anticipating a stratagem that will be relevant later, we note
that the invariant ansatz \eqref{plane1} is not the only way to produce an $S^3$-invariant 
plane. Indeed, the two equations defining $A$ might also transform in the two-dimensional
irreducible representation of $S^3$, \ie, $A$ might be given by $\{x_1-x_2,x_2-x_3\}$. 
This eliminates any free parameters in $A$, while bringing back those in $B$ and $C$ to 
$16-1=15$, and the number of equations to $21$. In the present case, allowing the 
equations to transform non-trivially under the symmetry group does not uncover any 
new solutions. In later examples it will.

\subsection{\texorpdfstring{$\zet_2\times\zet_2$}{Z2xZ2}-invariant conics}

The next case of interest is the subgroup $\zet_2\times\zet_2\subset S^5$,
with generators acting by exchanging $(x_1,x_3)$ and $(x_2,x_4)$ respectively.
\footnote{$\zet_2\times\zet_2$-invariant conics at $\psi=0$ have also been 
studied in \cite{mustata}.}
Assuming the equations for the plane to be invariant leads to the ansatz
\begin{equation}
\eqlabel{plane2}
A : \left\{ \begin{array}{c}
x_1+x_3 + a_1 x_5\, \\
x_2+x_4 + a_2 x_5 
\end{array}\right\}
\end{equation}
We here see immediately that for any value of $a_1,a_2$, the plane \eqref{plane2} contains,
at $x_5=0$, one of the $375$ isolated lines discussed in the previous section.
Therefore, if the quintic has any conic in such a plane, the residual cubic in eq.\ 
\eqref{practice} will be reducible, so that there are then, actually, two conics in that plane.
A priori, we do not expect to find any such conic at all, since the quartic curve 
residual to the line in the plane would need to develop four nodes where we only have 
two parameters at our disposal to move the plane. The symmetries help, however, as we 
shall see presently.

We solve the equations in \eqref{plane2} for $x_1$ and $x_2$, and make the ansatz
\begin{equation}
\eqlabel{conic2}
b_6 x_3^2+ b_5 x_3 x_4 + b_4 x_3 x_5 + b_3 x_4^2 + b_2 x_4 x_5+ b_1 x_5^2
\end{equation}
for the equation defining the conic. This is invariant under $x_3\to x_1=-x_3-a_1 x_5$
and under $x_4\to x_2= - x_4- a_2 x_5$, precisely if
\begin{equation}
\eqlabel{impose}
b_5=0\,,\qquad b_2=a_2 b_3\,,\qquad b_4 = a_1 b_6
\end{equation}
Eliminating the cubic, we find that the solution of \eqref{practice} is, in
the gauge $b_6=1$, reduced to the three equations
\begin{equation}
\eqlabel{Z2Z2invariant}
\begin{split}
1-a_1^5-a_2^5+5 a_1^3 b_1-5 a_1 b_1^2 &=0\\
a_2^3-\psi b_1-a_1^3 b_3+2 a_1 b_1 b_3 &=0 \\
a_2-\psi b_3+a_1 b_3^2 &=0
\end{split}
\end{equation}
for the four variables $a_1,a_2,b_1,b_3$. We see that this describes
two one-parameter families of conics for each $\psi$: the first equation admits 
two solutions for $b_1$, the middle equation then determines $b_3$ uniquely, while 
the third equation relates $a_2$ and $a_1$. The two families share the planes, but
not any conics. As in the previous subsection, the eq.\ \eqref{Z2Z2invariant} describes 
only an open patch of the families. Below, we will see a bit of the compactification, 
as dictated by the embedding in the moduli space of conics in $\projective^4$. As an 
example, one might verify the symmetry under exchange of $a_1$ and $a_2$.

\subsection{Taking advantage of phase symmetries}

Going slowly enough over the previous discussion reveals an option for finding
further solutions: not all of the $\zet_2\times\zet_2$ symmetry group under which 
the plane \eqref{plane2} is invariant need to fix the two conics in that plane
individually. Instead, the two conics might be exchanged by one generator,
and fixed by the other. In particular, we can choose the diagonal 
$\zet_2^+\subset\zet_2\times\zet_2$ to fix the two conics, and the exchange 
of $(x_1,x_3)$ to exchange them. The generator of $\zet_2^+$ acts as
\begin{equation}
\eqlabel{Z2p}
(x_1,x_2,x_3,x_4,x_5)\mapsto (x_3,x_4,x_1,x_2,x_5)
\end{equation}
This relaxes the constraint \eqref{impose} to
\begin{equation}
\eqlabel{relax}
b_2= a_2b_3 + \frac 12 a_1 b_5 \,,\qquad\qquad
b_4 = a_1 b_6 + \frac 12 a_2 b_5
\end{equation}
To simplify the equations further, we employ a device that will be useful also later: 
note that the ansatz \eqref{plane2} is covariant under a $(\zet_5)^2$ subgroup of the 
group of phase symmetries, provided we act in a particular
way on the coefficients of conic and cubic. So we may absorb those phase symmetries by
appropriate variable substitutions, and thereby reduce the degree of the equations. In
the present case, we first solve for $b_1,b_2,b_3,b_4$, and $a_2$ linearly. (Doing this
excludes the families above, on which the rank of the equations is reduced.)
Still working in the gauge $b_6=1$, we then substitute
\begin{equation}
\psi = \tilde \psi a_1 
\end{equation}
and the remaining equations depend only on $a_1^5$. Introducing $\tilde a_1=a_1^5$ as a new 
variable reduces the degree sufficiently to be able to fully understand
the equations. Indeed, $\tilde a_1$ appears only linearly, and the remaining relations
for $b_5$ and $\tilde \psi$ boil down to
\begin{equation}
-64 \tilde\psi+2 \tilde\psi^6-32 b_5^2+
11 \tilde\psi^5 b_5^2+
25 \tilde\psi^4 b_5^4+
30 \tilde\psi^3 b_5^6+
20 \tilde\psi^2 b_5^8+
7 \tilde\psi b_5^{10}+
b_5^{12} = p_1 p_2 p_3
\end{equation}
with
\begin{equation}
\eqlabel{split}
\begin{split}
p_1 &= -2+\tilde\psi+b_5^2 \\
p_2 &=
16+8 \tilde\psi+4 \tilde\psi^2+2 \tilde\psi^3+\tilde\psi^4+
(8+8 \tilde\psi+6 \tilde\psi^2+4 \tilde\psi^3) b_5^2 +
\\&\qquad\qquad\qquad\qquad\qquad\qquad\qquad\qquad
+(4+6 \tilde\psi+6 \tilde\psi^2) b_5^4+
(2+4 \tilde\psi) b_5^6+
b_5^8 \\
p_3 &=2 \tilde\psi+b_5^2
\end{split}
\end{equation}
We may then substitute back $a_1$ and $\psi$, to obtain the relations in a way that will be
useful later on. As an example, we write the equations corresponding to $p_1$:
\begin{equation}
\eqlabel{p1}
\begin{split}
64+5 \psi^2 a_1^3-40 \psi a_1^4+12 a_1^5 &=0 \\
\psi-2 a_1+a_1 b_5^2 &=0
\end{split}
\end{equation}
The two solutions for $b_5$ correspond, as it should be, to the two conics that exist in the
plane determined by the first equation.
The remaining parameters of our ansatz are given by
\begin{equation}
\eqlabel{remaining1}
\begin{array}{c}
a_2 = a_1\,, \qquad 
b_1 = \frac{1}{8} (-\psi a_1+6 a_1^2+2 a_1^2 b_5 )\,, \\[.2cm]
b_2 = a_1+\frac12 a_1 b_5 \,,\qquad
b_3 = 1 \,,\qquad
b_4 = a_1+ \frac12 a_1 b_5 \,,\qquad
b_6 = 1
\end{array}
\end{equation}
For future reference, we note that the conics corresponding to \eqref{p1} are, in 
addition to $\zet_2^+$, invariant under the group $\zet_2^-$ whose generator acts as
\begin{equation}
\eqlabel{Z2m}
\zet_2^-: (x_1,x_2,x_3,x_4,x_5)\mapsto (x_2,x_1,x_4,x_3,x_5)
\end{equation}
Indeed, when $a_2=a_1$, the equations for the plane \eqref{plane2} are exchanged under
that $\zet_2^-$, while the conic \eqref{conic2} is invariant as $b_6=b_3$ and
$b_4=b_2$.

For $p_3$, the relations analogous to \eqref{p1} are:
\begin{equation}
\eqlabel{p3}
\begin{split}
\psi^{10}+4096 a_1^5-160 \psi^5 a_1^5+1024 a_1^{10} &=0 \\
2 \psi+a_1 b_5^2 &=0
\end{split}
\end{equation}
and the explicit solution is
\begin{equation}
\eqlabel{remaining3}
\begin{array}{c}
\displaystyle
a_2 = \frac{\psi^2}{4 a_1} \,,\qquad
b_1 = - \frac{\psi^5}{64 a_1^3} + \frac{a_1^2}{2} + \frac{1}{16}\psi^2 b_5\\[.2cm]
\displaystyle
b_2 = -\frac{\psi^3}{8 a_1^2}+ \frac{1}{2} a_1 b_5\,,\qquad
b_3 = -\frac{\psi}{2a_1}\,,\qquad
b_4 = a_1+\frac{\psi^2 b_5}{8 a_1} \,,\qquad
b_6 =1
\end{array}
\end{equation}
These solutions are invariant only under $\zet_2^+$. 

Anticipating some of the later discussion, we note that the conic \eqref{conic2} is 
reducible when the $b_i$'s take the values in \eqref{remaining3}, provided $b_5$ satisfies 
the condition in \eqref{p3}. So we see that in fact, the plane \eqref{plane2}
with $4 a_2 a_1=\psi^2$ and $a_1$ satisfying the first equation in \eqref{p3} meets the 
quintic in a collection of five lines.

We have not discussed in detail the conics corresponding to the factor $p_2$ in 
\eqref{split}. In fact, that solution arises from \eqref{p1}, \eqref{remaining1} simply 
by the phase symmetry acting on $x_2$ and $x_4$. For instance, one may check that 
instead of $a_2=a_1$, merely $a_2^5=a_1^5$ holds on that solution. 

For completeness, we take a brief look at solutions with $b_6=b_3=0$ (which was excluded
by our above choice of gauge). We find that the only such solutions are
\begin{equation}
\eqlabel{completeness}
a_1=a_2= b_2=b_3=b_4=b_6=0 \,,\qquad\qquad
b_5^2 = 5 \psi b_1^2
\end{equation}
so that we recover the conics studied in \cite{mowa}.

\subsection{\texorpdfstring{$\zet_2^-$}{Z2}-invariant conics}

An important feature of the previous subsection was that the group $\zet_2^+$ (see 
eq.\ \eqref{Z2p}) acts with unit determinant on $(x_1,x_2,x_3,x_4,x_5)$, and the number 
of equations matched the number of variables also in the reduced problem. We were
able to fully reduce those equations, and thereby isolate the components of 
$\calh_{\rm conic}$ invariant under $\zet_2^+$.

As an ultimate possibility, we now study conics that are invariant under the action
\begin{equation}
\eqlabel{finalsym}
\zet_2^-: (x_1,x_2,x_3,x_4,x_5)\mapsto (x_2,x_1,x_4,x_3,x_5)
\end{equation}
As symmetries of $\projective^4$, the groups $\zet_2^+$ and $\zet_2^-$ are of course 
equivalent. But, as it turns out, we get a new class of solutions if we modify the ansatz 
for the plane $A$ containing the conic, and take one equation to be invariant, and
the other to transform with a sign (this being equivalent to the way \eqref{Z2m} acted on
\eqref{plane2}):
\begin{equation}
\eqlabel{plane3}
A:\left\{ 
\begin{array}{c}
a_1(x_1+x_2)+a_2(x_3+x_4) + x_5 \\
(x_1-x_2) + a_3 (x_3-x_4) 
\end{array}
\right\}
\end{equation}
The equation for the conic has to be invariant (we eliminate only $x_5$ in order
to make the symmetry manifest):
\begin{equation}
\eqlabel{conic3}
B: \{ b_1 (x_1+x_2)^2 + b_2 (x_1+x_2)(x_3+x_4) + b_3 (x_3+x_4)^2 
+ b_4 (x_3-x_4)^2 \}
\end{equation}
Now absorb the phase symmetries by substituting
\begin{equation}
\eqlabel{substitute}
\begin{array}{c}
a_1\to a_1^{-1/5}\,,\qquad
a_2\to a_2^{-1/5}\,, \qquad
a_3\to a_3\, a_1^{1/5} a_2^{-1/5}\,,\qquad
\psi\to\psi\, a_1^{-2/5}a_2^{-2/5} \\[.5cm]
b_1\to b_1\, a_1^{-2/5}\,,\qquad
b_2\to b_2\, a_1^{-1/5} a_2^{-1/5} \,,\qquad
b_3\to b_3\, a_2^{-2/5} \,,\qquad
b_4\to b_4\, a_2^{-2/5}
\end{array}
\end{equation}
where we denote the new variables by the same letters as the old ones.
Then, in the patch $b_2=1$, the $a_1,a_2,a_3^2,b_4$ can be solved for linearly.
When substituted back, we remain with two equations of relatively high degree
involving $b_1, b_3$, and $\psi$, 
\begin{equation}
\eqlabel{equations}
q_1(b_1,b_3,\psi) = q_2(b_1,b_3,\psi) =0 
\end{equation}
These can be further reduced if we exploit the inherent symmetry of \eqref{conic3}
exchanging 
$b_1$ and $b_3$, and substitute $b_1-b_3=u$, $b_1b_3=v$. Then, we eliminate 
$\psi$ by computing the resultant of those two equations, to decompose
the set of conics invariant under the $\zet_2^-$-symmetry \eqref{finalsym} as much
as possible into constituents:
\begin{equation}
\eqlabel{resultant}
{\rm Resultant}(q_1,q_2;\psi)(u,v) \propto  (9-12u+16v)\cdot (1-4v) \cdot (u^2-4v)
\cdot Q_m
\end{equation}
Here we have excluded factors that do not lead to a solution of the original system,
because the equations for $a_1,a_2,a_3,b_4$ that we solved earlier actually became 
singular. To each factor of \eqref{resultant}, there corresponds a component%
\footnote{We are not claiming here that all of those components are irreducible.
It's just the best we can do at this point.} 
of $\calh_{\rm conics}$ that can be reconstructed in the following way: given 
a pair $(u,v)$ for which that factor vanishes, we find a $\psi$ solving 
the equations \eqref{equations} (the existence of a common root of $q_1$ and
$q_2$ being precisely the characterization of the 
resultant), and then a unique set of $a_1,a_2,a_3^2,b_4$ solving the equations 
for a conic on the quintic under the $\zet_2^-$-invariant ansatz \eqref{plane3}, 
\eqref{conic3}, after absorbing the $\zet_5\times\zet_5$ phase symmetries as in
\eqref{substitute}. Undoing that substitution introduces 2 fifth roots of unity,
one of which corresponds to the phase of $\psi$ that originally parameterized
the family of quintics, while the other is a genuine label of a conic contained therein.
We thus obtain various branches of conics for each factor of the resultant 
\eqref{resultant}, depending on which solution we choose. These branches will interact
in various ways as (the original) $\psi$ (appearing in \eqref{WW}) is varied around
the moduli space. The different factors of \eqref{resultant} might split further
under this procedure (but they will not mix). An obvious splitting arises when
we remember that the map $(b_1,b_3)\to (u,v)$ is actually two-to-one. For example,
\begin{equation}
\eqlabel{realized}
9-12u+16 v = (3-4b_1)(3-4b_3)
\end{equation}
so that the first factor in \eqref{resultant} actually describes two sets of
conics in that sense. Also, although we did not bother pointing this out, it is 
clear that the equations are invariant under $a_3\to-a_3$, so we also need to
choose a sign for $a_3$ when we go back.

The main component of $\calh_{\rm conics}$ that we 
found is characterized by the factor,
\begin{multline}
\eqlabel{monster}
Q_m = \\
\begin{array}[t]{l} 
\scriptscriptstyle
-140544+1312896 u-6157536 u^2+20560128 u^3-55739073 u^4+126082635 u^5-240562314 u^6+ 389983296 u^7
-517794816 u^8\\[-.3cm]
\scriptscriptstyle
+526312752 u^9-386382096 u^{10}+195989568 u^{11}-64755264 u^{12} +12180480 u^{13}-890112 u^{14}
+3526016 v-35327360 u v\\[-.3cm]
\scriptscriptstyle
+164085512 u^2 v-490389848 u^3 v +1119877362 u^4 v-2054126078 u^5 v+2822178044 u^6 v
-2674914608 u^7 v+1703155648 u^8 v\\[-.3cm]
\scriptscriptstyle
- 783769296 u^9 v+331207776 u^{10} v-160872256 u^{11} v+74273920 u^{12} v-20820224 u^{13} v
+1915392 u^{14} v-12887824 v^2\\[-.3cm]
\scriptscriptstyle
+156888240 u v^2-794842896 u^2 v^2+1924669488 u^3 v^2-1861954446 u^4 v^2-560979783 u^5 v^2
+2742716878 u^6 v^2\\[-.3cm]
\scriptscriptstyle
-2532259552 u^7 v^2+1357646032 u^8 v^2-522831968 u^9 v^2+146600816 u^{10} v^2-62851072 u^{11} v^2
+15982784 u^{12} v^2\\[-.3cm]
\scriptscriptstyle
+5083904 u^{13} v^2-966912 u^{14} v^2-110365024 v^3+1349538976 u v^3-5573477584 u^2 v^3
+9890213496 u^3 v^3-8559117395 u^4 v^3\\[-.3cm]
\scriptscriptstyle
+4959540898 u^5 v^3-3410214400 u^6 v^3+1529015152 u^7 v^3
+33207472 u^8 v^3-335159488 u^9 v^3+398796352 u^{10} v^3\\[-.3cm]
\scriptscriptstyle
-199492096 u^{11} v^3+27038976 u^{12} v^3+206336 u^{13} v^3-181248 u^{14} v^3-1058031072 v^4
+7328123536 u v^4-16155350056 u^2 v^4\\[-.3cm]
\scriptscriptstyle
+17717423024 u^3 v^4-18250232092 u^4 v^4
+21436831296 u^5 v^4-16619578848 u^6 v^4+8429844448 u^7 v^4-3758257792 u^8 v^4\\[-.3cm]
\scriptscriptstyle
+1108256896 u^9 v^4-254268672 u^{10} v^4+129069056 u^{11} v^4-52310016 u^{12} v^4
+4696064 u^{13} v^4+135168 u^{14} v^4-1271515824 v^5\\[-.3cm]
\scriptscriptstyle
-3515100512 u v^5+23558245664 u^2 v^5-33532680832 u^3 v^5+15994006832 u^4 v^5+1748284832 u^5 v^5
-6786182656 u^6 v^5\\[-.3cm]
\scriptscriptstyle
+5719888128 u^7 v^5-2496033024 u^8 v^5+544198656 u^9 v^5-76355584 u^{10} v^5
+61018112 u^{11} v^5+3694592 u^{12} v^5\\[-.3cm]
\scriptscriptstyle
-1777664 u^{13} v^5+8485369664 v^6-39975494784 u v^6+76393384256 u^2 v^6
-82428927744 u^3 v^6+64625199040 u^4 v^6\\[-.3cm]
\scriptscriptstyle
-46320419072 u^5 v^6+28977470976 u^6 v^6
-13398732800 u^7 v^6+5270946816 u^8 v^6-1532405760 u^9 v^6+165781504 u^{10} v^6\\[-.3cm]
\scriptscriptstyle
-21012480 u^{11} v^6+2473984 u^{12} v^6+3324777728 v^7
-4295229696 u v^7-21543773440 u^2 v^7+57614347264 u^3 v^7-56924167424 u^4 v^7\\[-.3cm]
\scriptscriptstyle
+37388443136 u^5 v^7-22467149824 u^6 v^7+7877509120 u^7 v^7-2088861696 u^8 v^7+451772416 u^9 v^7
-78036992 u^{10} v^7\\[-.3cm]
\scriptscriptstyle
+6029312 u^{11} v^7-8267872256 v^8+34670743552 u v^8-42630860800 u^2 v^8+7651102720 u^3 v^8
+12061375488 u^4 v^8
-7382695936 u^5 v^8\\[-.3cm]
\scriptscriptstyle
+6715727872 u^6 v^8-2138890240 u^7 v^8+730972160 u^8 v^8-79429632 u^9 v^8+8498114560 v^9
-35617423360 u v^9+49886576640 u^2 v^9\\[-.3cm]
\scriptscriptstyle
-22319595520 u^3 v^9-1981624320 u^4 v^9+2571264000 u^5 v^9
-2668953600 u^6 v^9+344719360 u^7 v^9-2792865792 v^{10}\\[-.3cm]
\scriptscriptstyle
+10876387328 u v^{10}-12904677376 u^2 v^{10}
+2578120704 u^3 v^{10}+3163045888 u^4 v^{10}-473956352 u^5 v^{10}+320798720 v^{11}\\[-.3cm]
\scriptscriptstyle
-891617280 u v^{11}+732364800 u^2 v^{11}-87818240 u^3 v^{11}
\end{array}
\end{multline}

\subsection{Relationships}

We now describe how the solutions of \eqref{practice} that we have found so far
by imposing certain symmetries fit together as components of the Hilbert scheme, 
$\calh_{\rm conics}$, of conics on the one-parameter family of quintics \eqref{WW}. 
(As emphasized before, we do not claim that we have identified all components, 
nor that all components that we have found are irreducible.)

\subsubsection{$\zet_2\times\zet_2$ meets $S^3$}

First of all, we point out that the family of $S^3$-invariant conics found by Musta\c t\v a 
and the family of $\zet_2\times\zet_2$-invariant conics \eqref{Z2Z2invariant} meet.
A common member occurs in the first family if we put $a_1=1$, $a_2=0$ in 
\eqref{plane1}, where the conic acquires some additional symmetry, and in particular 
the $\zet_2\times\zet_2$ symmetry manifest in \eqref{plane2}. (Note that this solves 
\eqref{S3invariant} and that the conic \eqref{conic1} is also invariant because $b_1=1$.) 
To, conversely, exhibit that conic on the $\zet_2\times\zet_2$-invariant family,
we first need to go to a different patch of the moduli space. We note that the plane 
in \eqref{plane2} is equivalent to 
\begin{equation}
\eqlabel{equivalent}
\left\{\begin{array}{c}
x_1+x_3+ \tilde a_1(x_2+x_4) \\
\tilde a_2(x_2+x_4) + x_5 
\end{array}
\right\}
\end{equation}
where $\tilde a_1=-a_1/a_2,\tilde a_2=a_2^{-1}$. If we now put $\tilde a_1=1,\tilde
a_2=0$, we recover the plane invariant under both $S^3$ and $\zet_2\times\zet_2$ symmetry 
that we just discussed. The quintic meets that plane in a conic plus three lines. 
One of those is the line common to all the planes, the irreducible conic is the
$S^4$-enhancement in one of the families of $\zet_2\times\zet_2$-invariant
conics, while the remaining two lines represent the second family.

Another example of a common member of the two types of families 
occurs when $\tilde a_1=0, \tilde a_2=1$ in \eqref{equivalent}. This conic appears on 
the family invariant under permutation of $(x_2,x_4,x_5)$ in the limit $a_2=a_1$, $a_1^{-1}=0$ 
in the appropriate version of the plane \eqref{plane1}.

It is a useful exercise to write the equations for the families of $\zet_2\times\zet_2$-invariant 
conics in the patch with coordinates $\tilde a_1,\tilde a_2$. The parameterization
\eqref{conic2} was valid as long as we could eliminate $x_1$ and $x_2$---this is not 
possible when $\tilde a_2\to 0$ in \eqref{equivalent}. So eliminating $x_5$ instead of 
$x_2$, we write for the conic
\begin{equation}
\tilde b_1 x_2^2+\tilde a_1 \tilde b_3 x_2 x_3+
\tilde b_3 x_3^2+\tilde b_4 x_2 x_4+\tilde a_1 \tilde b_3 x_3 x_4+
\tilde b_1 x_4^2
\end{equation}
Then, in the patch $\tilde b_1=1$, the equations for the family are
\begin{equation}
\eqlabel{transform}
\begin{split}
-5 \tilde a_1+5 \tilde a_1^3 \tilde b_3+(1-\tilde a_1^5-\tilde a_2^5) \tilde b_3^2 &=0 \\
-1-4 \tilde a_1^5-4 \tilde a_2^5+(1+4 \tilde a_1^5+4 \tilde a_2^5) \tilde b_4
+(1-\tilde a_1^5-\tilde a_2^5) \tilde b_4^2 &=0 \\
-10 \tilde a_1^3-5 \psi \tilde a_2+5 \tilde a_1^3 \tilde b_4+
\tilde b_3 (1+4 \tilde a_1^5+4 \tilde a_2^5)+
2\tilde b_3\tilde b_4(1- \tilde a_1^5 -\tilde a_2^5) &=0
\end{split}
\end{equation}
These equations have a structure comparable to that of \eqref{Z2Z2invariant},
and in the relevant open patches the two systems are completely equivalent.

\subsubsection{Reducible conics}

Secondly, we record that the $\zet_2\times\zet_2$-invariant families contain reducible
conics. We have seen this in the discussion we just had at $\tilde a_1=1$, $\tilde a_2=0$.
Another example is $\tilde a_1=\tilde a_2=0$, which plane, $x_1+x_3=x_5=0$, meets 
the quintic in the union of five of the $375$ isolated lines that we discussed in 
the previous section. Yet another example is $a_1=1$, $a_2=0$ in \eqref{plane2}.

\subsubsection{$\zet_2^-$ meets $\zet_2^+$.}

Thirdly, we study in more detail the locus $u^2-4v=(b_1-b_3)^2=0$ corresponding to 
the vanishing of the third factor of \eqref{resultant} in the set of $\zet_2^-$-invariant
conics. We find that the relevant planes have $a_2=a_1$, $a_3=-1$ in \eqref{plane3},
and after eliminating the cubic, the resulting equations split in two components. 
One solution may be written as
\begin{equation}
\eqlabel{remaining4}
a_1=a_2=\frac 1\psi \,,\qquad a_3=-1\,,\qquad
b_2=1\,,\qquad b_3=b_1\,,\qquad b_4=-1+2b_1
\end{equation}
with $b_1$ satisfying
\begin{equation}
\eqlabel{p4}
-16+\psi^5+64 b_1+\psi^5 b_1-64 b_1^2+4 \psi^5 b_1^2=0
\end{equation}
The other is characterized by the vanishing of
\begin{equation}
\eqlabel{p5}
3-20 \psi a_1+5 \psi^2 a_1^2+512 a_1^5
\end{equation}
After $a_1\to  1/(2a_1)$, we may recognize this equation as being equivalent to the first
line in \eqref{p1}. Indeed, when $a_2=a_1$, and $a_3=-1$, the ansatz \eqref{plane3}
is also invariant under $\zet_2^+$ from \eqref{Z2p}, and with appropriate 
substitutions, the solution given by \eqref{p5} is nothing but that 
in \eqref{p1}, \eqref{remaining1}. One can also check that the solution \eqref{remaining4}, 
\eqref{p4} is contained in the family \eqref{Z2Z2invariant} at $a_1=a_2=\psi/2$. 

\subsubsection{A family of reducible conics}

Next, we discuss the conics associated with the vanishing of the second factor in 
\eqref{resultant}, 
\begin{equation}
\eqlabel{lach}
1-4v=1-4 b_1b_3
\end{equation} 
We see quite rapidly that under this condition, the conic \eqref{conic3} 
is reducible (remember that we work with $b_2=1$). The
two components must be lines on the mirror quintic, which are completely understood as
reviewed in the previous section. Since it is clear that the relevant lines are not on 
the list of $375$ isolated lines, they must belong to the families containing
the van Geemen lines. This indeed makes sense: for generic $\psi$, we have two 
distinct families of lines in $Y_\psi$ that meet in a curve $K_\psi \subset Y_\psi$. 
Each point in $K_\psi$ is the intersection point of two lines, one from each
family, which together can be properly viewed as a reducible conic. (The part of
the Clemens conjecture stating that rational curves are generically disjoint
obviously does not hold here.) In other words, each $Y_\psi$ contains a family of
reducible conics parameterized by $K_\psi$. At certain isolated points in $K_\psi$, the
reducible conic acquires the $\zet_2^-$ symmetry we have discussed, and shows
up on our list. 

As a further check, one may start from the $\zet_2^-$-invariant solutions with
$1-4v=0$ and verify that it indeed deforms as a one-parameter family of reducible 
conics, which generically break the $\zet_2^-$ symmetry. 

Moreover, we can now remember the $\zet_2^+$-invariant solution  \eqref{p3}, 
\eqref{remaining3}, which conics were also reducible with components not on the 
list of $375$. These must also belong to the family of reducible conics parameterized
by $K_\psi$. In fact, they must correspond to singular points on $K_\psi$ since
we actually have four such lines (\ie, two reducible conics) in the corresponding 
planes.

\subsubsection{Another coincidence}

Finally, we note that the conics associated with the vanishing of the first factor in 
\eqref{resultant}, which according to \eqref{realized} can be realized for instance at 
$b_1=3/4$, actually belong to the Musta\c t\v a family of $S^3$-invariant conics. This
can easily be checked.

\section{Summary So Far}
\label{sofar}

We have seen that the generic $Y_\psi$ contains (at least) three types of families 
of conics, and we have identified a number of isolated conics. We have not attempted
to enumerate the solutions, mostly because we did not work out the global description 
of all the families. Numerical methods indicate that these might in fact be all 
solutions: elementary search algorithms (such as those of Mathematica) return only 
solutions on one of our families, or isolated solutions with a non-trivial symmetry 
group. (This has to be taken with a dose of skepsis, because such algorithms have a
higher chance of finding solutions with symmetry or those occuring in families.)
We have also checked that there are no solutions with other types of symmetry 
enhancements than those we have discussed.

There is first of all the family with $S^3$ symmetry, parameterized by solutions of
\eqref{S3invariant}, and originally found by Musta\c t\v a \cite{mustata}. Secondly, 
we found two families of conics invariant under the action of a $\zet_2\times\zet_2$
symmetry group. These are parameterized by solutions of \eqref{Z2Z2invariant}, and 
have the interesting feature of sharing their planes. The $\zet_2\times\zet_2$ 
invariant families meet the $S^3$-invariant families in various ways, and contain 
reducible conics. Thirdly, there is a family consisting entirely of reducible conics. 
This family can be obtained by intersecting the van Geemen families of lines. We have 
not written down the equations describing that family globally, but identified 
two members, one with $\zet_2^+$ symmetry \eqref{p3} and one with $\zet_2^-$ symmetry 
\eqref{lach}.

Turning to the isolated conics, we have first of all those studied in \cite{mowa},
see eq.\ \eqref{completeness}. Secondly, we have the $\zet_2^+\times\zet_2^-$-invariant
conics of \eqref{p1}, \eqref{remaining1} (their $\zet_5$-orbit was also 
discussed around there). One may check (for instance numerically), that these
conics are in fact isolated also in the space of conics without any discrete
symmetries. Finally, we have $\zet_2^-$-invariant conics associated with solutions
of \eqref{monster}, which are also isolated forgetting the symmetry. As it stands, 
\eqref{monster} is not particular useful of course. It is somewhat unfortunate
that we have not been able to make further progress on those equations, for instance
with the purpose of checking whether the corresponding part of $\calh_{\rm conics}$
is in fact irreducible or not. In the next section, we will retreat to studying 
the expansion in the large complex structure limit.

We now switch to the main topic of interest in this paper, namely the Abel-Jacobi
image of $\calh_{\rm conics}$. We first of all dispose of the families:
recall that the isolated lines have a vanishing Abel-Jacobi image (in the sense 
described in the introduction, and in section \ref{vangeemen}), and the van Geemen 
lines give rise to the inhomogeneity \eqref{vginh}. As described above, the families 
are all algebraically equivalent to some combination of those lines, and therefore
they do not give rise to any new inhomogeneity.

We only remain with the isolated conics. We shall denote the family of conics 
(component of $\calh_{\rm conics}$) studied in \cite{mowa}, $\cali_0$. That
associated with \eqref{p1} will be called ``the first component'', $\cali_1$, 
and that of \eqref{monster}, (which might still be reducible), the ``main
component'', $\cali_2$.

\section{Expansion in Large Complex Structure Limit}
\label{largecomplex}

We realized in the previous section that the expression \eqref{monster} is too
large to allow writing down explicitly all coefficients determining the ``main 
component'', $\cali_2$, of the space of $\zet_2^-$-invariant conics, at least not 
without significantly increased computing power. Progress is still possible, however.

\subsection{Newton-Puiseux expansion}
\label{newtonpuiseux}

The main idea is easy to describe: instead of reducing the equations satisfied 
by the parameters in \eqref{plane3}, \eqref{conic3} algebraically, we expand 
those parameters in (fractional) power series around large complex structure point 
$\psi=\infty$, and determine the expansion coefficients recursively from the equations. 
This is in principle sufficient to calculate the expansion of $\calw(z)$ for the purpose 
of testing mirror symmetry. In practice, the calculation is limited to the first few 
orders in the expansion. The method will also not allow to easily calculate monodromies 
around complex structure moduli space, which would be desirable in order to fix the 
solution of the homogeneous Picard-Fuchs equation in $\calw$.

For what we'll call the ``first component'' of $\calh_{\rm conics}$, $\cali_1$, we are 
able to calculate the inhomogeneity exactly, see \eqref{ffirst}. This allows expansion 
to much higher order, calculation of monodromies, and is also a useful cross-check on the 
calculations around $\cali_2$. 

But let's first be a bit more general and take $v$ as any one of the parameters entering 
an ansatz for the curve $\calc\subset \caly$ under consideration (\eg, one of the $a_i$ 
or $b_j$ in \eqref{plane3}, \eqref{conic3}). (We could also imagine $\caly$ to be a more 
general family of algebraic varieties than the mirror quintic, degenerating in some way,
and $\calc$ to be some general algebraic cycle.) For a one-parameter family, 
$v$ will, as a function of $\psi$, satisfy a parameter-dependent polynomial equation 
\begin{equation}
\eqlabel{Pvpsi}
P(v,\psi)=0
\end{equation}
(obtained in the example by projecting eq.\ \eqref{practice} onto $(v,\psi)$, and 
generally at least as complicated as \eqref{monster}).  Let's assume that $P$ is
irreducible.

Following Newton, we can study the behavior of the roots of \eqref{Pvpsi} as 
$\psi\to\infty$, by looking at the polygon spanned by monomials with non-zero coefficients 
in $P(v,\psi)$: say
\begin{equation}
\eqlabel{say}
P(v,\psi) = \sum_{m,n} p_{m,n} v^m \psi^n\,,
\end{equation}
and let $\Pi$ be the convex hull of points $(m,n)\in\zet^2$ with $p_{m,n}\neq 0$.
Because we are looking at $\psi\to\infty$, the interesting part of $\Pi$ actually 
is its {\it upper boundary}, $\widehat\Pi$, which is the set of point $(m,n)\in\Pi$ 
such that $(m,n')\notin \Pi$ for $n'>n$. This $\widehat\Pi$ consists of a finite
sequence of segments of decreasing slope and varying length.

Let $v_k(\psi)$ (with $1\le k\le E$) be one of the $E$ roots of $P(v,\psi)=0$ for fixed 
$\psi$. (Here, $E:={\max}\{m,p_{m,n}\neq 0 \text{ for some $n$} \}$ is the $v$-degree 
of $P$.). There is then a rational number $\alpha_k$ such that
\begin{equation}
c_0 := \lim_{\psi\to\infty} \psi^{-\alpha_k} v_k(\psi) \notin \{ 0, \infty \}
\end{equation}
In other words, we are making an ansatz of the form
\begin{equation}
\eqlabel{newtonansatz}
v_k(\psi) = \psi^{\alpha_k} \bigl( c_0 + \calo(\psi^{-\beta}) \bigr) \quad \text{with 
$c_0\neq 0$ and $\beta>0$}
\end{equation}
plug this into \eqref{say}, and collect terms of same order in $\psi$:
\begin{equation}
\eqlabel{collect}
P(v_k(\psi),\psi) = \psi^{\alpha_{P}} \bigl( P_0(c_0) + \calo(\psi^{-\beta})\bigr)
\end{equation}
Then,
\begin{equation}
\eqlabel{alphaP}
\alpha_{P} = \max\{\alpha_k m+ n, p_{m,n}\neq 0 \}
\end{equation}
and
\begin{equation}
\eqlabel{P0}
P_0(c_0) = \sum_{\alpha_k m+n = \alpha_{P}} p_{m,n} c_0^m
\end{equation}
For $P_0(c_0)=0$ to have a non-zero solution, there must be at least two
non-zero terms in $P_0(c_0)$. This shows that (i) $\alpha_k$ must be the 
negative slope of one of the upper edges of $\Pi$ (\ie, of a segment of 
$\widehat\Pi$), and (ii) $P_0(c_0)$ is the sum of $p_{m,n}c_0^m$ along that
segment. In particular, the degree of $P_0(c_0)$ is the length of (the
projection onto the $m$-axis of) that segment.

Having found the lowest order term, we can proceed with the expansion 
\eqref{newtonansatz} to higher order. Several things can happen: for example,
the polynomial $P_0(c_0)$ might be reducible, or the sub-leading terms in 
\eqref{collect} might vanish together with $P_0(c_0)$. The next term in 
\eqref{collect} that does not vanish after imposing the leading order equation
will determine the exponent $\beta$ in \eqref{newtonansatz}. Past this, and 
if the original $P$ is irreducible, we are set to calculate the coefficients 
in 
\begin{equation}
\eqlabel{fractional}
v_k(\psi) = \psi^{\alpha_k} \sum_{d=0}^\infty c_d \psi^{-\beta d}
\end{equation}
recursively from the $\psi$-expansion \eqref{collect}. It is in the nature
of things that the coefficients $c_d$ for $d>1$ are finite algebraic expressions
in $c_0$, modulo $P_0(c_0)=0$.

In a somewhat more formal language, we can view $v$ as generator of an algebraic 
extension of the field of rational functions on $\psi$-space, of degree $E$. 
When localizing that extension at $\psi=\infty$, it splits into extensions of the 
local field of power (Laurent) series in $\psi^{-1}$, of degree $e_k$ given by the length 
of the corresponding segment of $\widehat\Pi$. (Observe that $\sum e_k = E$.)
The generators of these local extensions are precisely the Puiseux series 
\eqref{fractional}. Geometrically, we think of an algebraic curve \eqref{Pvpsi} as 
an $E$-fold cover of $\psi$-space, and what we are doing is simply parameterizing 
the various branches at $\psi=\infty$.

An underlying piece of structure is hidden in the following fact: the coefficients
$p_{m,n}$ are, generally speaking, algebraic combinations of the coefficients 
entering the definition of $\caly$ (and, possibly, the ansatz for $\calc$). In 
other words, the $p_{m,n}$ live in the field $K$ over which the underlying algebraic 
variety (and, possibly, the ansatz for $\calc$) is defined. As we have seen above,
to any branch of \eqref{Pvpsi} is associated a polynomial $P_0\in K[c_0]$ whose 
vanishing determines the leading coefficient $c_0$, and all other coefficients are 
algebraic combinations of $c_0$. Thus, each branch belongs to a certain algebraic 
extension $K(c_0)$ of the residue field $K$ at $\psi=\infty$. 

Given this structure, we now think of the branches of solutions of \eqref{Pvpsi} with
the same leading exponent and extension of residue field (splitting $P_0$ into irreducible 
factors if necessary) together in one group. It should be kept in mind, however, that
$P_0$ does not fully characterize the local extension because as discussed
above, sub-leading terms in $P(v,\psi)$ could play a significant role. And the
global extension determined by $P(v,\psi)$ itself of course is not visible in the local
expansion of any given group.

When we do not have the power to reduce to an equation of the type \eqref{Pvpsi}
explicitly, we can still obtain the Puiseux expansion of the various parameters by 
studying the original larger system of equations \eqref{practice}. Assuming the
underlying family to be one-dimensional, the discussion is similar, with Newton polygon 
replaced by Newton polyhedron, and vectors of leading exponents. The comments about 
importance of sub-leading terms in the expansion of the equations however become more 
acute. Indeed, with more equations, there are more ways in which they can be degenerate, 
and the actual extension of $K$ governing each group of solutions might only be determined 
at higher order in the expansion.

There are several more possibilities with higher-dimensional families: for example,
a family of cycles could define a transcendental extension of the parameter space, 
or require an additional blowup on top of the extension. 

We also mention an interesting alternative point of view on the expansion we have 
discussed: so far, we started with curves embedded in $\projective^4$, and imagined 
solving the equations that determined which of those would lie in the quintic 
hypersurface, then passing to the limit $\psi\to\infty$. Instead, we might also first
go to large complex structure, and note that the mirror quintic degenerates there 
into the union of 5 copies of $\projective^3$. While curves on the quintic are 
virtually rigid, curves in projective space have a large number of parameters. So the 
problem to study is the lifting of these moduli spaces under the perturbation away 
from large complex structure. This description is more intrinsic and presumably better
suited to understand the mirror symmetry.

Finally, we note that in our example, the initial family $\caly$ is defined of course
over $K=\rationals$. In fact, it is not entirely clear how to imagine families of 
Calabi-Yau manifolds defined intrinsically over a number field other than the rationals, 
other than by specializing parameters of a higher-dimensional family.\footnote{I thank
Ron Donagi for a helpful discussion on this issue.} With a D-brane 
(in the form of an algebraic cycle) on top of $\caly$, extensions of $\rationals$ are 
forced on us.

\subsection{The first component}

Let's warm-up to Newton-Puiseux expansions on the first component $\cali_1$ of 
$\calh_{\rm conics}$, for which we can write down the equations globally in 
$\psi$. We begin by writing \eqref{p1} in the form
\begin{align}
\eqlabel{curveid1}
64+5 a^3 \psi^2 -40 a^4 \psi+12 a^5 &=0\\
\eqlabel{curveid2}
-128 -5 a^2 \psi^3+40 a^3 \psi^2-12 a^4 \psi+64 b^2 &=0\,,
\end{align}
with the abbreviation $a\equiv a_1=a_2$, $b\equiv b_5$. Recall again that the vanishing 
of the first equation selects a plane \eqref{plane2} intersecting the quintic in (a 
line and) two conics determined by \eqref{conic2}, \eqref{remaining1}, and distinguished 
by the choice of root in \eqref{curveid2}.

\begin{figure}[th]
\begin{center}
\psfrag{1}{$1$}
\psfrag{2}{$2$}
\psfrag{3}{$3$}
\psfrag{4}{$4$}
\psfrag{5}{$5$}
\psfrag{a}{$m$}
\psfrag{psi}{$n$}
\epsfig{file=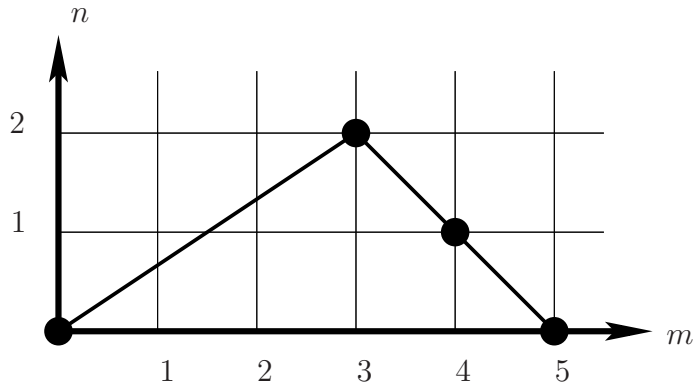,height=5cm}
\end{center}
\caption{Newton polygon of the equation \eqref{curveid1} for the first component of Hilbert 
scheme of conics on the mirror quintic. A full circle means that the coefficient of the
corresponding monomial $a^m\psi^n$ does not vanish.}
\label{newton}
\end{figure}

The Newton polygon of \eqref{curveid1} is shown in Fig.\ \ref{newton}.

We see that in the limit $\psi\to\infty$, the $5$ branches of solutions of 
\eqref{curveid1} split into 2 groups, with asymptotic exponent for $a$ given by the 
negative slope of the two upper segments of the Newton polygon: the first group has 
asymptotic behavior
\begin{equation}
a =  \psi^{-2/3} a_0 + \cdots
\end{equation}
with
\begin{equation}
64 + 5 a_0^3 = 0
\end{equation}
while the second,
\begin{equation}
a = \psi a_0 + \cdots
\end{equation}
with $a_0$ one of the roots of the equation
\begin{equation}
\eqlabel{under}
12 a_0^2 -40 a_0 + 5 = 0
\end{equation}
To determine the sub-leading terms, and the expansion of $b$, we plug
the leading order solution back into \eqref{curveid1}, \eqref{curveid2}.
We find that for the first group
\begin{equation}
b = b_0\psi^{5/6} +  \cdots
\end{equation}
with
\begin{equation}
64 b_0^2-5 a_0^2 = 0
\end{equation}
and the local expansion parameter is $\psi^{-5/3}$. 

For the second group, with local parameter $\psi^{-5}$, the leading order terms 
in \eqref{curveid2} vanish under the condition \eqref{under}. Therefore, we have 
to determine the leading order term in
\begin{equation}
b = b_0 + \cdots
\end{equation} 
from the sub-leading terms in \eqref{curveid2}:
\begin{equation}
64+5 a_0 a_1-60 a_0^2 a_1+24 a_0^3 a_1-32 b_0^2 =0
\end{equation}
Since this involves the coefficient $a_1$ in the expansion of $a$:
\begin{equation}
a = a_0 \psi + a_1 \psi^{-4} + \cdots
\end{equation}
we first have to first solve \eqref{curveid1} to that order. We find
\begin{equation}
a_1 = \frac{128(2729-852 a_0)}{425}
\end{equation}
and then the equation
\begin{equation}
\eqlabel{additional}
30-12 a_0+5 b_0^2 = 0
\end{equation}
In terms of the residue field at $\psi=\infty$, this is a second quadratic extension 
on top of \eqref{under}. 

To write the expansion in a more compact form that we will use later on, we introduce 
the more convenient
\begin{equation}
w = \frac{1}{5\psi} = z^{1/5}
\end{equation}
Then, on the first group of branches of \eqref{curveid1}, \eqref{curveid2},
we have
\begin{equation}
\eqlabel{expansion1}
\begin{split}
a&=\textstyle
-\frac{4\lambda^2}{5} w^{2/3}+\frac{128\lambda^4}{15} w^{7/3}-\frac{492800}{3} w^4+
\frac{534732800\lambda^2}{81} w^{17/3}-\frac{73034301440\lambda^4}{243} w^{22/3}+
\cdots \\
b&=\textstyle
\frac{\lambda^5}{1250} w^{-5/6}+\frac{14\lambda}{3} w^{5/6}-\frac{1132\lambda^3}{15}w^{5/2}
+\frac{1035344\lambda^5}{405} w^{25/6}-\frac{16285375600\lambda}{243} w^{35/6}+ \cdots
\end{split}
\end{equation}
with $\lambda$ one of the roots of the equation,
\begin{equation}
\eqlabel{oneof}
\lambda^6 = 5^4
\end{equation}
We might record here a typical feature of these expansions: the equation 
\eqref{oneof} means that $\lambda$ is, up to a root of unity, equal to $5^{2/3}$. 
A third root of unity is equivalent to the phase of the local expansion
parameter, $w^{5/3}$. The additional choice of sign is associated with the
choice of root in \eqref{curveid2}. The local monodromy $w\to\ee^{2\pi\ii}w$ 
permutes those branches cyclically. (As one might expect, the better local variable
is actually $z=w^5$. The concomitant $5$-th roots of unity in $a$ will cancel
out in \eqref{ffirst}.)

For the second group,
\begin{equation}
\eqlabel{expansion2}
\begin{split}
a &=\textstyle \frac{6+\lambda^2}{12} w^{-1}
-\frac{3200(-599+355 \lambda^2)}{17} w^4
+\frac{76800000(-23778566+14088349 \lambda^2)}{289} w^9- \cdots \\
b&=\textstyle \lambda+\frac{2000 (-74708 \lambda+44263 \lambda^3)}{51} w^5
-\frac{1600000 (-326587981456 \lambda+193497180065 \lambda^3)}{2601} w^{10} +\cdots
\end{split}
\end{equation}
where $\lambda$ is one of the roots of
\begin{equation}
\eqlabel{twoof}
5\lambda^4+20\lambda^2-48=0
\end{equation}
Note that while again the choice of sign for $\lambda$ originates from \eqref{curveid2},
the local monodromy $w\to\ee^{2\pi\ii}w$ acts trivially.

Before leaving this family of cycles for a while, we show the result of the
computation of the inhomogeneous Picard-Fuchs equation. The algorithm of 
\cite{mowa,newissues} can be applied without much change. The main complication
is that one has to keep the parameters $a$ and $b$ implicit throughout. Since
the line residual to the two conics has a vanishing superpotential (see section 
\ref{vangeemen}), the inhomogeneity should be odd under $b\to-b$. With standard 
conventions, such as reviewed in section \ref{vangeemen}, we find the Picard-Fuchs 
inhomogeneity associated with conics in $\cali_1$ to be:
\begin{equation}
\eqlabel{ffirst}
\begin{array}[t]{l} 
\displaystyle 
\qquad\qquad\call\int^C\Omega = f(z) \\[.5cm]
f(z) = \frac{1}{4\pi^2} \,
\frac{b}{320\,(-128 + 3 \psi^5)^3\, (-5308416 + 26104832\psi^5 + 459\psi^{10})^3}
\; \cdot\\
\scriptscriptstyle
\cdot[-3529208202219460015329116160 - 
5917959309462446377556508672\, a^4 \psi - 
24080174251679112693326807040\, a^3 \psi^2 
\\[-.3cm]\scriptscriptstyle
- 37102979749413690361774080000\, a^2\psi^3 + 
5322140674208202106664386560\, a\psi^4 
\\[-.3cm]\scriptscriptstyle
+ 377013614277474642973792665600\,\psi^5 
- 223673316478788106348117622784\, a^4\psi^6 + 
231620425022730366652294103040\, a^3\psi^7 
\\[-.3cm]\scriptscriptstyle
+ 577173365083785450174157946880\, a^2\psi^8 + 
1161971462867073400022583214080\, a\psi^9 
\\[-.3cm]\scriptscriptstyle
+ 1138625829170016488325937889280\,\psi^{10} 
- 162426814061060730487566237696\, a^4\psi^{11} + 
462200036747394287493017763840\, a^3\psi^{12}
\\[-.3cm]\scriptscriptstyle 
+ 196861662250863298084696227840\, a^2\psi^{13} - 
198567289143941889876285194240\, a\psi^{14} 
\\[-.3cm]\scriptscriptstyle
+ 
385678957625260010043531591680\,\psi^{15} 
- 188475902674373195063233609728\, a^4\psi^{16} + 
397300557436660139725013647360\, a^3\psi^{17} 
\\[-.3cm]\scriptscriptstyle
+ 468813519263945326185655828480\, a^2\psi^{18} + 
479723528675140620247262822400\, a\psi^{19} 
\\[-.3cm]\scriptscriptstyle
+ 352752475928491530510768537600\,\psi^{20} 
- 39263076586488037778065981440\, a^4\psi^{21} + 
110777498597321397283848192000\, a^3\psi^{22} 
\\[-.3cm]\scriptscriptstyle
+ 42233632645599612642734899200\, a^2\psi^{23} + 
16695932913990986817444249600\, a\psi^{24} 
\\[-.3cm]\scriptscriptstyle
+ 5506564481958675778539356160\,\psi^{25} 
- 279092702543449176793939968\, a^4\psi^{26} + 
884770078321237750123069440\, a^3\psi^{27} 
\\[-.3cm]\scriptscriptstyle
+ 34251597272406042397900800\, a^2\psi^{28} - 
12180273406238980319477760\, a\psi^{29} 
\\[-.3cm]\scriptscriptstyle
- 7891860706457745044275200\,\psi^{30} 
+557447463014026659692544\, a^4\psi^{31} - 
1763923787950883886858240\, a^3\psi^{32} 
\\[-.3cm]\scriptscriptstyle
- 71223763050638247444480\, a^2\psi^{33} + 
4711857482247092305920\, a\psi^{34} 
\\[-.3cm]\scriptscriptstyle
+ 1639504965244195307520\,\psi^{35} 
- 34139433836832735744\, a^4\psi^{36} + 
110844573279392655360\, a^3\psi^{37} 
\\[-.3cm]\scriptscriptstyle
- 4645064401757907840\, a^2\psi^{38} - 
375748813003714560\, a\psi^{39} 
\\[-.3cm]\scriptscriptstyle
- 14770116391956480\,\psi^{40} 
+ 66315921005988\, a^4\psi^{41} - 
220588897640760\, a^3\psi^{42} + 
26084392488495 a^2 \psi^{43} + 
193405158000 a\psi^{44} ]
\end{array}
\end{equation}
We will not say much here about the structure of that result, just as we skipped the
detailed discussion of the geometry of $\cali_1$. Note however that the factor 
$-128+3\psi^5$ in the denominator indicates an interesting interaction 
of $\cali_1$ with the van Geemen lines (\cf, eq.\ \eqref{vginh}). It is easy to check
that the conics in $\cali_1$ become reducible there (although not only there). The 
other factor in the denominator is the discriminant of \eqref{curveid1}. 

In the expansion \eqref{expansion1}, \eqref{ffirst} becomes
\begin{equation}
\eqlabel{firstgroup}
4\pi^2 f(z) = \textstyle
\frac{25\lambda}{54} z^{1/6}-\frac{2003 \lambda^3}{8} z^{1/2}
+\frac{18846875\lambda^5}{486} z^{5/6}-
\frac{6020738135875 \lambda}{2187} z^{7/6}+\cdots
\end{equation}
and on the second group of branches, \eqref{expansion2}, we have
\begin{equation}
\eqlabel{secondgroup}
4\pi^2 f(z) = \textstyle
10000 (-7624 \lambda+4517 \lambda^3) z-\frac{4000000
(-520331498984 \lambda+308286536785 \lambda^3)}{51} z^2 + \cdots
\end{equation}

\subsection{The main component}

We now turn to Puiseux expansions of the solutions of \eqref{practice}
that satisfy $Q_m=0$ (see eq.\ \eqref{monster}). The ansatzs for plane and conic
are (see eqs.\ \eqref{plane3}, \eqref{conic3}), 
\begin{equation}
\begin{split}
A&:\left\{ 
\begin{array}{c}
a_1(x_1+x_2)+a_2(x_3+x_4) + x_5  \\
(x_1-x_2) + a_3 (x_3-x_4) 
\end{array}
\right\}
\\
B&: \{ b_1 (x_1+x_2)^2 + b_2 (x_1+x_2)(x_3+x_4) + b_3 (x_3+x_4)^2 
+ b_4 (x_3-x_4)^2 \}
\end{split}
\end{equation}
For completeness, we display the ansatz for the cubic, 
\begin{multline}
C:\{c_1 (x_1+x_2)^3 + c_2 (x_1+x_2)^2(x_3+x_4) + c_3(x_1+x_2)(x_3+x_4)^2
+c_4(x_3+x_4)^3 \\+ c_5 (x_1+x_2) (x_3-x_4)^2 + c_6 (x_3+x_4)(x_3-x_4)^2\}
\end{multline}
as well as the full set of relations,
\begin{equation}
\eqlabel{fullset}
\begin{array}{lcl}
\frac1{80}-\frac{a_1^5}5 - b_1c_1&\qquad\qquad 
& -\frac{\psi a_1}{16}+\frac{a_3^2}{8}-b_4c_1-b_1c_5\\
-a_1^4 a_2-b_2 c_1-b_1c_2 & &-\frac{\psi a_2}{16}- b_4c_2-b_2c_5-b_1c_6\\
\frac{\psi a_1}{16} -2 a_1^3 a_2^2-b_3c_1-b_2c_2-b_1 c_3 &&
-\frac{\psi a_1 a_3^2}{16}-b_4c_3-b_3c_5-b_2c_6\\
\frac{\psi a_2}{16} - 2a_1^2 a_2^3-b_3c_2-b_2c_3-b_1 c_4 & &
\frac{1}{8}-\frac{\psi a_2a_3^2}{16}-b_4 c_4-b_3c_6\\
-a_1a_2^4-b_3c_3-b_2c_4 & &
\frac{\psi a_1a_3^2}{16}+\frac{a_3^4}{16}-b_4c_5\\
\frac1{80}-\frac{a_2^5}5-b_3c_4 & &
\frac{1}{16}+\frac{\psi a_2a_3^2}{16}-b_4 c_6
\end{array}
\end{equation}
After scaling one of the $b_j$'s to $1$, we have $12$ equations for $12$
variables, in addition to $\psi$, which we want to turn into a local expansion 
parameter. For each of our variables $v_i$ ($i=1,\ldots, 12$), we make an ansatz
of the form
\begin{equation}
v_i = \sum_{d=0} (v_i)_d \psi^{\alpha_i - \beta d}
\end{equation}
with rational $\alpha_i$, $\beta$, plug into those equations, and solve order 
by order in $\psi$.\footnote{Implementing this requires more diligence and patience
than is appropriate to perhaps explain.}

Remembering the warnings emitted in subsection \ref{newtonpuiseux}, we have
a little bit of extra work to do at low order: the equations at lowest order
might not determine all $(v_i)_0$ immediately. They could also split into 
several pieces that lie on separate components of $\calh_{\rm conics}$.
For the latter issue, we keep only those that lie on $\cali_2$, \ie, which satisfy
\eqref{monster}. For the former, we continue to higher order. This determines
the local expansion parameter $\psi^{-\beta}$, and eventually, all obstructions are 
lifted, and we can mechanically solve the recursion. We identify (a power of) one 
of the $(v_i)_0$ as generator of the number field associated with the corresponding 
group of branches.
The information on the various groups belonging to $\cali_2$ is collected in 
Table \ref{groups}.

\begin{table}[ht]
\begin{tabular}{|l|l|l|l|}
\hline
 & exponents, in order        & local &  generator of number field and \\
\# & $(a_1,a_2,a_3,b_1,b_2,b_3,b_4)$ & par. & minimal polynomial \\ \hline
1 & $(-1,-1,0,0,0,0,0^*)$ & $\psi^{-5}$ & $\lambda= (a_3^{10})_0$\;;
\begin{tabular}{l}
$\scriptstyle \lambda^{10}-243 \lambda^9+27675 \lambda^8-1529140 \lambda^7$\\[-.3cm]
$\scriptstyle +49599473 \lambda^6+221079468 \lambda^5+49599473 \lambda^4$ \\[-.3cm]
$\scriptstyle -1529140 \lambda^3+27675 \lambda^2-243 \lambda+1$
\end{tabular}\\\hline
2 & $(0,0,-\frac12,1,1,\frac 12,0^*)$ & $\psi^{-1/2}$ & $\lambda=(b_3)_0$ \;; $\lambda^{10}-62208$ 
\\\hline
3 & $(\frac 17,0,-\frac 17,\frac 27,-\frac 47,-\frac 57,0^*)$ & $\psi^{-5/7}$ & 
$\lambda=(2a_3^{2}a_2)_0$\;; $\lambda^{14}-5\lambda^7+5$\\\hline
4 &$(0,\frac12,\frac12,-\frac12,0^*,\frac12,\frac12)$ & $\psi^{-5/2}$ & 
$\lambda=(32b_3^5)_0$\;; $\lambda^4+11\lambda^2-1$ \\\hline
\end{tabular}
\caption{Groups of branches of $\zet_2^-$-invariant conics. 
Some choices capture fairly obvious symmetries of the ansatz: $a_3\mapsto-a_3$ 
corresponds to exchange of $x_3$ and $x_4$. Multiplication of $(x_1,x_2)$ and $(x_3,x_4)$ 
by opposite fifth roots of unity can also be absorbed without touching the local
expansion parameter. The exchange $a_1\leftrightarrow a_2, a_3\leftrightarrow 1/a_3$ 
produces further groups, but leaves the first invariant (this is related to the
symmetry $\lambda\to 1/\lambda$). In each group, $0^{*}$ is the 
exponent of the variable that we have found convenient to scale to $1$.}
\label{groups}
\end{table}

To illustrate the complexity, we give some of the lowest order terms in the expansion of
the fourth group in the table:
\begin{equation}
\begin{split}
a_1 a_2 & = \textstyle
\frac{-8\lambda-\lambda^3}{20} \psi^{1/2}
+\frac{-47-4\lambda^2}{20} \psi^{-2}
+ \frac{130229\lambda+11743\lambda^3}{40} \psi^{-9/2} + \cdots \\
a_2^5 &= \textstyle
\frac{\lambda^3}{32} \psi^{5/2} + \frac{3+2\lambda^2}{32} + 
\frac{-1105\lambda-101\lambda^3}{64} \psi^{-5/2} +\cdots\\
a_3^2/a_2^4 &= \textstyle
- \frac{16(21+2\lambda^2)}{15} \psi^{-1} + \frac{32(5024\lambda
+453\lambda^3)}{15} \psi^{-7/2} - \frac{16(71099+6411\lambda^2)}{3} \psi^{-6}
+\cdots
\end{split}
\end{equation}
We have also ventured into the calculation of the inhomogeneous Picard-Fuchs
equation for these cycles. Working order by order in the residue calculus of
\cite{mowa}, we obtain for the third group in Table \ref{groups}:
\begin{multline}
\eqlabel{solving}
f(z) = \textstyle\frac{\ii\lambda^{1/2}}{4\pi^2}
\Bigl[ 
\frac{25(5\lambda^5-\lambda^{12})}{343} z^{1/7}
+\frac{500(94-17\lambda^7)}{2401} z^{2/7}
+\frac{225(70585\lambda^2-31748\lambda^9)}{16807} z^{3/7}\\
\textstyle
+\frac{400(2394125\lambda^4-191028\lambda^{11})}{117649} z^{4/7}
+\frac{3875(245997065\lambda^6-63500311\lambda^{13})}{823543} z^{5/7}
+\cdots
\Bigr]
\end{multline}
(as usual, $z=(5\psi)^{-5}$).
This illustrates again the general structure we have 
been discussing: the seventh root of unity is the phase of the local 
expansion parameter $z^{1/7}$. The additional square-root in \eqref{solving} originates 
from the choice of sign of $a_3$ in \eqref{fullset}: the exchange of $x_3$ and $x_4$ 
changes the cycle class by a sign. The remaining irrationality is intrinsic to the 
group of algebraic cycles under consideration.

\section{Expansion in Large Volume Limit}
\label{largevolume}

We are now ready to study the A-model expansion of the space-time superpotential.
The main focus is the so-called multi-cover formula that relates the A-model
expansion to the BPS content of the supersymmetric space-time theory.

Schematically, the general prediction of ref.\ \cite{oova} was that a single BPS state 
of charge $\beta$ should make a contribution to the space-time superpotential of the 
form
\begin{equation}
\eqlabel{oovast}
\calw_\beta(q) \sim {\rm Li}_2(q^\beta) \sim \sum_k \frac{q^{\beta k}}{k^2}
\end{equation}
where $t=\log q$ is the complex scalar in the supermultiplet coupling to $\beta$, 
and ${\rm Li}_2$ is the standard Euler's di-logarithm function. The sum over $k$ 
originates as the Laplace transform of the D0-brane charge in the M-theory derivation 
of \eqref{oovast}. In the context of \cite{oova} one assumes a local A-model 
setup with a non-compact Lagrangian as D-brane, where $t$ represents K\"ahler
moduli as well as freely adjustable D-brane moduli.

If $n_\beta$ is the degeneracy of BPS states of charge $\beta$, the total
superpotential is
\begin{equation}
\eqlabel{degeneracy}
\calw = \sum_\beta n_\beta\calw_\beta \,.
\end{equation}
This superpotential, together with its higher-derivative generalizations in the
context of the open topological string, is equivalently computable from a sum over 
world-sheet instantons with boundary on the background D-brane. Disentangling 
the contributions in the various charge sectors, see, \eg, \cite{lmv}, leads to 
the customary relations between open Gromov-Witten invariants and BPS (Ooguri-Vafa) 
invariants.

For example, for the standard (``inner'') brane on the (resolved) conifold at zero 
framing, there are two BPS states of charge $(0,1)$ and $(1,-1)$, respectively, with
a space-time superpotential:
\begin{equation}
\eqlabel{conifold}
\calw (t,u) = \sum_{k=1}^\infty\Bigl( \frac{\ee^{k u}}{k^2} + \frac{\ee^{k(t-u)}}{k^2}\Bigr)
\end{equation}
for the K\"ahler modulus $t$, and the open string (D-brane) modulus $u$.
In the worldsheet computation of \eqref{conifold}, one counts holomorphic maps
from the disk to the conifold, with boundary mapping to the Lagrangian submanifold
wrapped by the D-brane. The sum over $k$ originates from those maps that factor via 
degree $k$ multi-coverings of the disk by itself, such as
\begin{equation}
\eqlabel{degreek}
z \mapsto z^k
\end{equation}

It has been noted in \cite{open,newissues,ahmm} that these multi-cover formulas
are not suitable in the context of compact manifolds. The main physical reason is that 
anomalies prevent a full separation of open and closed string moduli, while open 
Gromov-Witten invariants are not defined in general. The basic conundrum is already 
implicit in \cite{oova}, where the masses of 2-d BPS solitons are determined by the 
critical values of the superpotential, which however is only generated by integrating
out those very solitons. The issue could be resolved if it were possible to make 
sense of $\calw$ off-shell, \ie, away from its critical points, ideally without 
additional information from the K\"ahler potential. In non-compact situations, 
certain natural choices are suggested by the symmetries of the asymptotic 
geometry \cite{agva,akv}. In compact situations, one class of off-shell choices was 
proposed in \cite{joso,ahmm}, and a somewhat different one in \cite{ghkk}.

Our way to deal with the ambiguities is to consider the 
critical points of the superpotential in the $u$-direction (\cf, \eqref{critical}).
For the conifold,
\begin{equation}
\begin{split}
\del_u\calw(t,u) & = -\log (1-\ee^{u})  +\log(1-\ee^{t-u}) = 0 \\
\Rightarrow & \ee^u = \pm \ee^{t/2}
\end{split}
\end{equation}
Then the difference of critical values is given by
\begin{equation}
\eqlabel{elevate}
\calw(t,u_+)- \calw(t,u_-) = 4 \sum_{k\;{\rm odd}} \frac{\ee^{tk/2}}{k^2} 
\end{equation}
This on-shell superpotential encodes less information than \eqref{conifold}, but
depends on fewer choices. If the inner brane on the conifold as a local model
captures enough of the global geometry, one can elevate \eqref{elevate}
to a multi-cover formula instead of \eqref{oovast}.
\begin{equation}
\eqlabel{thisversion}
\calw (q) = \sum_{d\;{\rm odd}}\tilde n_d q^{d/2} =
\sum_{d,k\;{\rm odd}} n_d \frac{q^{dk/2}}{k^2} 
\end{equation}
Indeed, this modification of \eqref{oovast} was found in \cite{open} to relate the 
rational open Gromov-Witten invariants $\tilde n_d$ of the real quintic to integer 
invariants $n_d$, that fit into a larger framework of real enumerative geometry.

In \cite{newissues,ahmm}, other modifications of the di-logarithm were identified,
such as
\begin{equation}
\eqlabel{triple}
\sum_{3\nmid k} \frac{q^{d k/3}}{k^2}
\end{equation}
albeit without a description of either local or global A-model geometry.

Through the examples of the present paper, we will see that \eqref{thisversion} and 
\eqref{triple} are just the simplest versions of a much more elaborate class of 
``multi-cover'' formulas. The relevance of certain arithmetic functions in these new 
multi-cover formulas is rather intriguing, and indicative of deeper connections
between mirror symmetry and number theory that we hope to explore elsewhere.

\subsection{Van Geemen lines}

We first return to the van Geemen families of lines. Their inhomogeneous
Picard-Fuchs equation was calculated in section \ref{vangeemen},
\begin{equation}
\call\calw_B(z) = f_{\rm van\;Geemen}(z)
\end{equation}
where 
\begin{equation}
\call= \theta^4 - 5 z(5\theta+1)(5\theta+2)(5\theta+3)(5\theta+4)
\end{equation}
$\theta=\frac{d}{d\ln z}$, $z=(5\psi)^{-5}$, and, with $1+\omega+\omega^2=0$,
\begin{equation}
f_{\rm van\; Geemen}(z) = \frac{1+2\omega}{4\pi^2}\,\cdot\,
\frac{32}{45}\,\cdot\,
\frac{\frac{63}{\psi^{5}}+\frac{1824}{\psi^{10}}-\frac{512}{\psi^{15}}}
{\Bigl(1-\frac{128}{3\psi^5}\Bigr)^{5/2}}
\end{equation}
It is straightforward to solve this equation in a power series around $z=0$,
and apply the usual mirror map to obtain 
\begin{equation}
\eqlabel{scared}
\begin{split}
\widehat{\calw}_A(q)& =4\pi^2\calw_A(q) = \frac{4\pi^2\calw_B}{\varpi_0}(z(q))  \\
&=\sqrt{-3}\,\bigl(\textstyle 140000 q+\frac{11148100000}{3} q^2+
\frac{5015947794500000}{27} q^3+
\frac{330137902935872500000}{27} q^4 \\
&\textstyle+\frac{76015582693256843498840000}{81} q^5+
\frac{57929080529317310275946498060000}{729} q^6+\cdots \bigr)
\end{split}
\end{equation}
Instead of being scared away by the growth of the numerators of the expansion coefficients, 
let us look at the denominators. We define $\tilde n_d$ as the coefficient of $q^d$:
\begin{equation}
\eqlabel{firstexample}
\widehat{\calw}_A(q) = \sum_{d=1}^\infty \tilde n_d q^d
\end{equation}
Remarkably, the $\tilde n_d$ are not rational numbers, in distinction to all previous
examples in the literature. From expansion to large order, we observe that the denominator 
of $\tilde n_d$ grows as $3^d$, but otherwise contains at most a factor of $d^2$, \ie, we have
\begin{equation}
\eqlabel{discou}
d^2 3^d \frac{\tilde n_d}{\sqrt{-3}} \in \zet
\end{equation}
Given previous experience, in which the $d^2 \tilde n_d$ were always (rational) integers,
the result \eqref{discou} could seem a bit disappointing. On the other hand, the
denominators are remarkably smaller than those in $\calw_B$. Roughly speaking, the
mirror map reduces $(d!)^2$ to $d^2$. It is natural to expect that the factors of $d^2$ 
in the denominator can be removed by an appropriate multi-cover formula. It is remarkable 
that such a formula indeed exists!

On expanding
\begin{equation}
\eqlabel{firstD}
\widehat{\calw}_A(q) = \sum_{d=1}^\infty n_d \sum_{k=1}^\infty \frac{\chi(k)}{k^2} q^{dk}
\end{equation}
where $\chi(k)$ depends on the residue of $k\bmod 3$,
\begin{equation}
\eqlabel{dirichlet}
\chi(k) = \begin{cases} 0 & k\equiv 0\bmod 3 \\
1 & k\equiv 1\bmod 3\\
-1 &k\equiv 2\bmod 3
\end{cases}
\end{equation}
we find
\begin{equation}
3^d \frac{n_d}{\sqrt{-3}} \in \zet
\end{equation}
The first few $n_d$ are%
\footnote{The first convincing case is
$\tilde n_{11} 
= \sqrt{-3}\,\frac{5195025975738748330135719454410630564027766563867792882680000}{3^{10} 11^2}$,
compared with $n_{11} = \tilde n_{11}+\tilde n_1/11^2=
\sqrt{-3}\,\frac{42934098973047506860625780614963888958907161684857860740000}{3^{10}}$}
\begin{equation}
\begin{array}{rcl}
n_1 & = & \sqrt{-3}\, 140000 \\
n_2 & = & \sqrt{-3}\, \frac{11148205000}{3}\\
n_3 & = & \sqrt{-3}\,\frac{5015947794500000}{27}\\
n_4 & = & \sqrt{-3}\,\frac{330137902960955725000}{27} \\
\vdots & & \qquad\quad \vdots 
\end{array}
\end{equation}

The function $\chi(k)$ in \eqref{dirichlet} is, of course, just the quadratic
character modulo 3, which is the non-trivial Dirichlet character of order 3, 
one of the standard arithmetic functions of algebraic number theory.
\begin{equation}
\chi(k) = \left(\frac{k}{3}\right) 
\end{equation}
Incidentally, we may now recognize \eqref{thisversion} and \eqref{triple} as having
a rather similar form, with $\chi(k)$ replaced by the trivial (principal) 
Dirichlet character of order 2, and $3$, respectively.

\subsection{The D-logarithm}

The results so far motivate us to introduce more general twists of the di-logarithm,
which we will call the D-logarithm, of the form
\begin{equation}
\eqlabel{Dlog}
{\rm Li}_2^{\rm D}(x) = \sum_{k=1}^\infty \frac{a_k}{k^2} x^k
\end{equation}
where $(a_k)$ are sequences of numbers that we will specify (see subsections \ref{definition}
and \ref{redefinition}).

The purpose of the D-logarithm is to serve as a refinement of the multi-cover
formula \eqref{oovast} for
general D-brane superpotentials. The notation and terminology, however, is suggested
by the special case that $a_k=\chi(k)$ is a Dirichlet character,
and in which we write,
\begin{equation}
{\rm Li}_2^{(\chi)}(x) = \sum_{k=1}^\infty \frac{\chi(k)}{k^2} x^k
\end{equation}
When $\chi$ is a trivial Dirichlet character, we recover the formulas 
of \cite{open,newissues}, while $\chi(k)=\left(\frac{k}{3}\right)$ is relevant for
the van Geemen lines.

The original occurrence of twists of this type of course is in Dirichlet L-functions,
\begin{equation}
L(s;\chi) = \sum_{n=1}^\infty \frac{\chi(n)}{n^s}
\end{equation}
In fact, given the relation of special values
\begin{equation}
L(2;\chi) = {\rm Li}_2^{(\chi)}(1)
\end{equation}
one might view the D-logarithm as a natural alternative analytic continuation of
those special values. We will see below however that the coefficients $(a_k)$
relevant for the D-logarithm are in general different from those occuring in 
typical L-series.

\subsection{First component, first group}

So let us consider now the A-model expansion of the superpotential for the
family of conics that we have been calling $\cali_1$, and whose inhomogeneous
Picard-Fuchs equation is calculated in \eqref{ffirst}.

As discussed in section \ref{largecomplex}, the $10$ branches of $\cali_1$ over the
complex structure moduli space fall into two groups in the large volume limit. The 
first group contains $6$ branches that are in correspondence with the solutions of 
\eqref{oneof}; the second has $4$ branches and is governed by \eqref{twoof}. We 
do not have an A-model interpretation of this structure, but we can calculate the 
superpotential as a solution of \eqref{ffirst}. This will allow us to delineate the 
definition of the D-logarithm.
 

In the expansion \eqref{firstgroup}, we find after the standard mirror map:
\begin{equation}
\eqlabel{facilitate}
\begin{split}
\widehat{\calw}_A&(q) = \textstyle 
600\cdot 5^{2/3}\cdot q^{1/6}- 100150 q^{1/2}
+\frac{30155000}{3} \cdot 5^{1/3} \cdot q^{5/6}
-\frac{1965946252000}{1323}\cdot 5^{2/3}\cdot q^{7/6}\\
\textstyle
+ &\textstyle \frac{12016906931000}{9} q^{3/2}
-\frac{7939808112480350000}{29403}\cdot 5^{1/3}\cdot q^{11/6}
+\frac{21851198476716995185000}{369603} \cdot 5^{2/3}\cdot q^{13/6} \\
-&\textstyle\frac{205546823520516323768}{3} q^{5/2}
+\frac{94301250909743023365521125000}{5688387}\cdot 5^{1/3}\cdot q^{17/6}\\
-&\textstyle
\frac{795791304680793507631175999075000}{191850201} \cdot 5^{2/3}\cdot q^{19/6}
+ 
\frac{783657804098608936611454866250}{147} q^{7/2}
+\cdots
\end{split}
\end{equation}
We have here substituted $\lambda=5^{2/3}$ in order to facilitate the following
observations. Denoting the coefficient of $q^{d/6}$ by $\tilde n_d$, we 
have (all of what follows is confirmed to rather high order, up to $d \gtrsim 600$)
\nxt
when $2\mid d$, $\tilde n_d=0$
\nxt 
when $3\mid d$, $d^2\tilde n_d\in\zet$
\nxt
when $(d,6)=1$, $d^2 3^{\lfloor\frac{3d}{4}\rfloor}\tilde n_d\in 5^{\frac{2}{3}}\zet$
when $d=1\bmod 3$ and $\in 5^{\frac{1}{3}}\zet$ when $d=2\bmod 3$.

\medskip

\noindent
There are some similarities, but also noticeable differences to \eqref{scared}: 
\bnxt
The irrationality of the $\tilde n_d$ is not just an overall factor for $\widehat\calw_A$.
\bnxt
The denominators of the irrational $\tilde n_d$ grow with $d$ as a power of $3$, although
slightly less rapidly than \eqref{scared}. 
\bnxt The $\tilde n_d$ for $3\mid d$ are integer up to a factor of $d^2$.

\medskip

\noindent
As for the van Geemen lines \eqref{firstD}, our goal now is to describe the D-logarithm 
such that via
\begin{equation}
\eqlabel{want}
\widehat\calw_A(q) = \sum_d\tilde n_d q^{d/6} = \sum_{d} n_d {\rm Li}_2^{\rm D}(q^{d/6})
= \sum_{d} n_d \sum_k \frac{a_k}{k^2} q^{dk/6}
\end{equation}
the $n_d$ will have the remaining $d^2$ dropped from their denominators.
Some important points are clear at this stage already. 
\anxt The $n_d$ will remain irrational, and we are prepared to live with a growing denominator. 
\anxt The $a_k$ appearing in the D-logarithm will also be irrational. The most natural way
to capture the symmetries of \eqref{oneof} is that the $a_k$ are rational up to a factor of
$5^{2d(k-1)/3}$. 
\anxt This means in particular that the $a_k$ must depend on $d$. Given the symmetries,
one expects that the $a_k$ will depend on $d\bmod 3$.

\medskip

Before looking for a solution to these constraints, we add the following piece of
information: in a putative physics interpretation, to be further discussed below,
the $n_d$ should be related to a
degeneracy of appropriate BPS states that multiplies those states' contribution to
the space-time superpotential, see eq.\ \eqref{degeneracy}. Looking back at
eqs.\ \eqref{thisversion}, \eqref{triple}, and \eqref{firstD}, we see that the
variations of the multi-cover formula are related to the vacuum structure of the associated
D-brane configuration, and more precisely to the action of the symmetry group
of the algebraic equation determining that structure, the equation being
$\varphi^2=1$, $\varphi^3=1$, and $\omega^2+\omega+1=0$ in those three cases,
respectively. Together with the present data, this indicates that the correct 
version of the Ooguri-Vafa formula \eqref{oovast} depends on the {\it arithmetic 
properties} of the ``BPS degeneracies'' $n_d$, and more specifically on the action 
of the Galois group of the relevant number field.

A first hint that the result \eqref{facilitate} fits into the general framework comes
from the part of the expansion with only rational coefficients. As remarked above, the
$\tilde n_{3d}$ are rational, with denominator $(3d)^2$. And with the simple twist by
the principal character of order 2,
\begin{equation}
\eqlabel{firsthint}
\sum_{d\;{\rm odd}} \tilde n_{3d} q^{d/2} = \sum_{d,k\;{\rm odd}} \frac{n_{3d}}{k^2} q^{dk/2}
\end{equation}
one finds that the $n_{3d}$ are indeed integer. This is a quite non-trivial
check that there is no simple mistake in \eqref{ffirst}.

The irrational part of \eqref{facilitate} is governed by the number field
$K=\rationals(5^{1/3})$, with Galois completion $L=\rationals(5^{1/3},\sqrt{-3})$.
A crucial observation that sets us onto the right track is the following. 
Whereas $\tilde n_7$ and $\tilde n_{11}$ do not have an obvious congruence with 
$\tilde n_1$ (it is difficult to 
ascertain any statements about $d=5$ or multiples thereof, because $5$ divides the 
discriminant of the number field), one finds that in the combination
\begin{equation}
\eqlabel{crucial}
\tilde n_{13} - \frac{\tilde n_1}{13^2} = 5^{2/3}\cdot 
\frac{129297032406609431200}{3^7}
\end{equation}
the $13^2$ in the denominator cancels out. So we learn that $a_{13}=1$. Invoking 
rudimentary knowledge of elementary
algebraic number theory, we understand that indeed $13$ is special with respect to
our number theoretic situation: it is a prime that splits completely in the
field extension $L/\rationals$. Actually, already in $K$,
\begin{equation}
\eqlabel{splitcompletely}
x^3 - 5 = (x+2)(x+5)(x+6) \bmod 13
\end{equation}
That \eqref{crucial} is not a coincidence can be confirmed by checking the next
primes with the same property: $67,127,\ldots$.

Now let's again pause and compare with the van Geemen lines: when $k=p$ is prime, 
$a_p=\left(\frac{p}{3}\right)=1$ when $p$ is a quadratic residue $\bmod 3$.
By reciprocity, this is the case precisely when $p$ splits completely in 
$\rationals(\sqrt{-3})$, \ie, $x^2+3=0$ has two solutions in $\zet/p$. This 
is clearly consistent with the observations around \eqref{splitcompletely}. 
When $p$ does not split in $\rationals(\sqrt{-3})$, $a_p=-1$. More generally 
we have the multiplicativity $a_{k_1k_2}=a_{k_1}a_{k_2}$.

Adding knowledge of all previous cases as well as \eqref{firsthint} gives us 
a very clear understanding of when $a_p=1$: either there is no field extension, or 
$p$ splits completely in the extension field. But how do we generalize the non-trivial
$\left(\frac{p}{3}\right)$ when $K=\rationals(5^{1/3})$, or some other 
number field? Since $L=K(\sqrt{-3})$ is a non-abelian extension, with Galois group $G=S^3$
the full permutation group, consultation of the literature might suggest
to try characters of higher-dimensional representations of $G$. This however
is incompatible with the expectation that the $a_p$ should themselves be general
irrationals, and seems in fact impossible to realize. Nevertheless, some
detective's work reveals that the correct answer indeed involves the structure 
of $G={\rm Gal}(L/\rationals)$, especially its action at the primes. We explain 
this next.

\subsection{D-logarithm \texorpdfstring{$\bmod k^2$}{mod k2}}
\label{definition}

In the previous subsections, we have listed various constraints that we expect
the D-logarithm to satisfy and collected hints that its underlying sequence
$(a_k)$ is determined by the arithmetic properties of the invariants $n_d$.
The for us characteristic property, namely that the formula \eqref{want} clear 
the $d^2$ from the denominators, can be used to determine the $a_k$ as 
algebraic numbers (integers) $\bmod k^2$. This is what we do here.%
\footnote{We give here a bottom-up presentation of the relevant results, 
following more or less the path along which we came to them. A more 
straight-forward mathematical definition will be written up elsewhere.}

Based on the cases involving Dirichlet characters, we believe that in general 
there are distinguished representatives for $a_k$, which then define the D-logarithm 
${\rm Li}_2^{\rm D}(x)$ as an analytic function of $x$. These distinguished
representatives should be obtained either from a proper physics derivation of
\eqref{oovast}, or an appropriate mathematical interpretation of the D-logarithm
as a multi-cover formula. 

We first make explicit that the D-logarithm depends on $n_d$ by rewriting \eqref{want} as
\begin{equation}
\eqlabel{want2}
\widehat\calw_A(q) = \sum_d\tilde n_d q^{d/6} = \sum_{d} n_d \sum_k 
\frac{a_k^{(d)}}{k^2} q^{dk/6}
\end{equation}
with the understanding that $a_k^{(d)}$ depends on the upper index $d \bmod 3$.
We know that $a_k^{(0)}=1$ for all odd $k$ and $a_p^{(d)}=1$ when $p$ is a prime that
splits completely in $K=\rationals(5^{1/3})$.%
\footnote{We will not make statements for $k$ that are not co-prime with the discriminant 
of the number field, which for our given $K$ is equal to $-3^3\cdot 5^2$.}

Plugging \eqref{facilitate} into \eqref{want2} we extract the following
values for $a_k^{(d)}\bmod k^2$ for $k=p$ the first few primes, $d=1,2$: (We are 
continuing to write $\lambda=5^{2/3}$ but of course other cube roots of $5^2$ would do 
as well.)
\begin{equation}
\begin{tabular}{c|c|c|}
$p$ & $a_p^{(1)} \bmod p^2$ & $a_p^{(2)} \bmod p^2$\\\hline
7 & 30 &  18 \\
11 & $82\cdot 5^{2/3}$ & $103 \cdot 5^{1/3}$ \\
13 & 1 & 1 \\
17 & $247 \cdot 5^{2/3}$ & 150 $\cdot 5^{1/3}$ \\
19 & 68 & 292 \\
23 & $59 \cdot 5^{2/3}$ & $477 \cdot 5^{1/3}$ \\
29 & $538 \cdot 5^{2/3}$ & $700\cdot 5^{1/3}$ \\
31 & 521 & 439 \\
\vdots & \vdots & \vdots
\end{tabular}
\end{equation}
The structure is fairly obvious: when $p\equiv 1\bmod 3$ (but doesn't split completely), 
$a_p^{(1)}$ and $a_p^{(2)}$ are the two roots $\bmod p^2$ of the equation
\begin{equation}
\eqlabel{cuberoot}
a^2 + a + 1 = 0
\end{equation}
When $p\equiv 2\bmod 3$, we find that $b^{(1)}= a_p^{(1)}/5^{2/3}$ is the root $\bmod p^2$ 
of the equation
\begin{equation}
\eqlabel{cr2}
25 b^3 = 1
\end{equation}
while $b^{(2)}=a_p^{(2)}/5^{1/3}$ is the root $\bmod p^2$ of the equation
\begin{equation}
\eqlabel{cr3}
5 b^3 = 1
\end{equation}
The solutions to \eqref{cr2} and \eqref{cr3} are unique, but how do we discriminate 
between the two roots of \eqref{cuberoot}? It turns out that the relevant information
comes from the Frobenius automorphism of the residue field extension at $p$: the
reduction of $a_p^{(d)}\bmod p$ should agree with $\bigl(5^{2d/3}\bigr)^{p-1}$,
\begin{equation}
a_p^{(d)} = \bigl(5^{2d/3}\bigr)^{p-1} \bmod p
\end{equation}
which then uniquely selects the solution of \eqref{cuberoot}. (For example $30=2=
(5^{2/3})^6=5^4\bmod 7$.)

This reduction also holds for $p\equiv 2\bmod 3$: 
\begin{equation}
\frac{a_p^{(1)}}{5^{2/3}} = \frac{\bigl(5^{2/3}\bigr)^{p-1}}{5^{2/3}} \bmod p
\end{equation}
and a similar equation holds for $d=2 \bmod 3$.

The distinction between $p=1$, or $2$, $\bmod3$ lies in the order of the Frobenius automorphism.
It is equal to $3$, or $2$, respectively, which corresponds exactly to the two non-trivial
cycle classes of the Galois group $G=S^3$.

To understand that these findings are in agreement with the conditions we
knew for $a_p^{(d)}=1$, we recall that when $p$ splits completely, or $n_d$ is rational,
the Frobenius automorphism acts trivially. This is true in the first situation because 
the local field extension is trivial, and in the latter can also be viewed as a 
consequence of Fermat's little theorem, $a^p=a\bmod p$ for any $a\in \zet$. This 
theorem also implies that the Frobenius acts on $n_d$ in the same fashion as on 
$5^{2d/3}$. Thus, we may summarize:
\begin{equation}
\eqlabel{pprime}
\text{
\framebox{\parbox{\textwidth-5.5cm}{
\centering{{\em For $p$ prime, $a_p^{(d)}$ is the lift $\bmod p^2$ of the 
Frobenius automorphism at $p$ acting on $n_d$.}}}}}
\end{equation}

What about $a_k$ when $k$ is {\it not} prime? A basic expectation is a multiplicative structure
relating $a_{k_1 k_2}$ to $a_{k_1}\cdot a_{k_2}$. It is clear however that this has 
to be refined because we have obtained $a_p$ only $\bmod p^2$. Moreover, the condition
that $a_k$ should be rational up to a factor $5^{2d(k-1)/3}$ (see page \pageref{want})
is not naively compatible with a multiplicative structure for $a_k^{(d)}$.

When $k=p_1p_2$ is the product of two distinct primes, one could imagine fixing representatives
of $a_{p_1}$ and $a_{p_2}$ and require that $a_{p_1p_2}$ coincide with $a_{p_1}a_{p_2}$ 
$\bmod (p_1p_2)^2$. But this is not sufficient when $k$ is divisible by a higher prime power 
$p^e$, $e>1$. To deal with this situation we introduce $e$ as an additional index and let
\begin{equation}
\eqlabel{lift}
\text{
\framebox{\parbox{\textwidth-4.5cm}{
\centering{$a_{p,e}^{(d)}$ be the lift $\bmod p^{2e}$ of the Frobenius at $p$ acting on $n_d$.}}}}
\end{equation}
For instance, when $p=1\bmod 3$, $a_{p,e}^{(d)}$ is the unique solution of \eqref{cuberoot}
$\bmod p^{2e}$ that agrees with the Frobenius $\bmod p$. We agree that $a_{p}^{(d)}=
a_{p,1}^{(d)}$.

The issue with the irrationality of the multiplicative structure can be resolved by 
mixing the two sequences $a_k^{(1)}$ and $a_k^{(2)}$.

Here then is the explicit algorithm that allows the recursive calculation of all $a_k^{(d)}
\bmod k^2$. For any prime $p$ dividing $k$, we denote by $e_p$ the largest power of
$p$ dividing $k$, and we require
\begin{equation}
\eqlabel{multiplicative}
\text{
\framebox{\parbox{\textwidth-10cm}{
\centering{$a_k^{(d)} = a_{p,e_p}^{(d k/p)} \cdot a_{k/p}^{(d)} \bmod p^{2e_p}$}}}}
\end{equation}
Imposing these conditions for all primes dividing $k$ determines $a_k^{(d)}$ uniquely
$\bmod k^2$.

An interesting aspect of eq.\ \eqref{multiplicative} is that the value of $a_k^{(d)}$ 
depends recursively on the representatives chosen for $a_{k/p}^{(d)}$ $\bmod k^2$ (which
is previously determined only $\bmod (k/p)^2$). The structure of \eqref{want2} however 
is such that those choices do not affect the integrality properties of the resulting 
$n_d$. This means that the representatives for $a_k^{(d)}\bmod k^2$ are not independent 
from one another, and is good evidence that more distinguished representatives should 
exist.

How this really works in detail is, of course, clearest in the examples. We have seen 
above that $a_7^{(1)}=30\bmod 7^2$, and easily find $a_{7,2}^{(1)}=1353\bmod 7^4$, so 
that $a_{49}^{(1)} = 40590$. Stripping the $\bmod$s and solving \eqref{want2}, we find
\begin{equation}
\begin{split}
n_1&=\tilde n_1 = 5^{2/3} \cdot 600\\
n_7 &= \tilde n_7 - \frac{30}{7^2} n_1 = -5^{2/3} \cdot\frac{40121362000}{3^3} \\
n_{49} &= \tilde n_{49} - \frac{30}{7^2} n_7  - \frac{40590}{7^4} n_1 =
5^{2/3} \cdot \frac{392867\cdots993000}{3^{33}}
\end{split}
\end{equation}
Plugging one into the other, we see that
\begin{equation}
\begin{split}
n_{49} &= \tilde n_{49} - \frac{a_7^{(1)}}{7^2} \tilde n_7 -
\Bigl(\frac{a_{49}^{(1)}}{7^4}- \frac{(a_{7}^{(1)})^2}{7^4}\Bigr) n_1 \qquad\qquad\qquad \\
&= \tilde n_{49} - \frac{a_7^{(1)}}{7^2} \tilde n_7 - 
\frac{\bigl(a_{7,2}^{(1)}- a_{7}^{(1)}\bigr)a_7^{(1)}}{7^4} n_1 \\
&= \tilde n_{49} - \frac{a_{7,2}^{(1)}}{7^2}\tilde n_7 
+\frac{a_{7,2}^{(1)}-a_7^{(1)}}{7^2}\Bigl(\tilde n_7-\frac{a_7^{(1)}}{7^2} n_1\Bigr)
\end{split}
\end{equation}
Since $a_{7,2}^{(1)}=a_7^{(1)}\bmod 7^2$, this shows that we may change $a_7^{(1)}$ by
multiples of $7^2$ without affecting the fact that the denominator of $n_{49}$ is not
divisible by $7$. 

We have checked up to $d\gtrsim 600$ that using the three formulas \eqref{pprime}, 
\eqref{lift}, and \eqref{multiplicative} in \eqref{want2} returns $n_d$ with no $d^2$
in the denominator.

\subsection{First component, second group}
\label{redefinition}

The definition of the D-logarithm $\bmod k^2$ that we have given in the previous
subsection was slanted towards the example \eqref{facilitate}. To see that the
prescriptions \eqref{pprime}, \eqref{lift}, and \eqref{multiplicative} make sense,
and are correct, in more generality, we here study the second group of branches of 
the first component $\cali_1$ of conics on the mirror quintic. 

The A-model expansion in this example begins
\begin{equation}
\eqlabel{thisexample}
\begin{split}
\widehat{\calw}_A(q) =&
10000 \lambda (-7624+4517 \lambda^2) q -\textstyle\frac{3200000\lambda 
(-40650831529+24084846092 \lambda^2)}{51} q^2  \\
&\textstyle\quad + \frac{250000 
\lambda (-5248611469517402890552+3109702672077500263451 \lambda^2}{3^3\cdot 17^2} q^3 \\
&\qquad\scriptstyle -\frac{2500000 \lambda (-781124731396525415521048504088+
462801576865994098449442008739 \lambda^2)}{3^3\cdot 17^3 }q^4 + \cdots
\end{split}
\end{equation}
where $\lambda$ is a root of
\begin{equation}
\eqlabel{minimal}
5\lambda^4+20 \lambda^2-48 = 0
\end{equation}
The Galois completion of $K=\rationals(\lambda)$ is $L=K(\sqrt{-3/5})$, with Galois group the 
dihedral group $D_4$. This can be visualized by arranging the four roots of \eqref{minimal} in a 
square,
\begin{equation}
\eqlabel{square}
\begin{array}{ccc}
\textstyle\sqrt{-2+2\sqrt{\frac{17}{5}}} & \line(1,0){40} &\sqrt{-2-2\sqrt{\frac{17}{5}}} \\
  \line(0,1){40} & & \line(0,1){40} \\
\textstyle-\sqrt{-2-2\sqrt{\frac{17}{5}}} & \line(1,0){40} &
-\sqrt{-2+2\sqrt{\frac{17}{5}}}
\end{array}
\end{equation}
The discriminant of $K$ is $-3\cdot 5^3\cdot 17^2$, and we may check that the denominators
in \eqref{thisexample} behave as $d^2 3^d 17^d$. We want to remove the $d^2$ using the 
appropriate D-logarithm.

The symmetries in this example do not completely constrain the irrationality of the 
$\tilde n_d$, which are general linear combinations of $\lambda$ and $\lambda^3$.
In other words, only the behavior of $\widehat\calw_A$ under $\pi$-rotation of 
\eqref{square} is fixed. As a consequence, the D-logarithm has a more severe
dependence on $d$ than in the previous example, leading to even more intricate checks
of the formalism.

So let us explain how \eqref{pprime} is implemented in the present example. We assume
that $n_d$ is an algebraic number with denominator vanishing at most at the discriminant
of $K$, and want to determine $a_p$ when $p$ does not divide the discriminant.
 
As one learns in algebraic number theory, the structure of $K/\rationals$ at the 
(rational) prime $p$ is related to the factorization $\bmod p$ of the minimal polynomial 
of an integral generator of $K$. In our examples, we could choose $\mu=5\lambda$, 
with minimal polynomial
\begin{equation}
\eqlabel{PPP}
P = \mu^4 + 100\mu^2 -6000
\end{equation}
Then to each factor of $P\bmod p$ is associated a prime (ideal) $\mathfrak{p}_i$ in $K$ ``lying
over  $p$'',
and the degree of that factor is the degree of the associated residue field extension
$(\zet_L/\mathfrak{p}_i)/(\zet/p)$. (As above, $L$ is the Galois completion of $K$, and
$\zet_L$ is the ring of integers in $L$.) This being an extension of a finite field, it has 
cyclic Galois group generated by a single element, called the Frobenius element $\sigma_i=
\sigma(\mathfrak{p_i}/p)$, which acts as $y\mapsto y^p$ in the residue field. In the Galois extension 
$L$, the Frobenius elements associated with different factors of $p$ would all be conjugate to 
each other, so determine a conjugacy class in the Galois group $G={\rm Gal}(L/\rationals)$.
In our case, $K/\rationals$ is not Galois, so we work with $\sigma_i$ that act as definite 
elements of $G$ on the roots of \eqref{PPP}. 

Now given $n_d$ as a (non-zero) algebraic number in $K$ we consider, for each $\mathfrak{p}_i$ 
dividing $p$, the algebraic number
\begin{equation}
z_i = \frac{\sigma_i(n_d)}{n_d}
\end{equation}
These $z_i$ themselves live in $L$, but not generally in $K$, and moreover, depend on $i$. 
So what do we mean in \eqref{pprime} by ``$a_p$ is given by the action of Frobenius at $p$''? 
The underlying idea, familiar in algebraic number theory, is to work ``locally around $p$'', \ie, 
approximate numbers $\bmod p$ (or, more generally, $\bmod p^e$ for $e>1$). In the local 
approximation, we can both find representatives for $z_i$, and interpolate between the 
different $z_i$, as $\mathfrak{p}_i$ varies over $p$. Moreover, given an approximation to 
order $p$, we can lift it $\bmod p^2$, and
this is our definition of $a_p$. The lifts $\bmod p^{2e}$, needed in \eqref{lift}, are then
obtained in a straightforward continuation of this procedure.

In formulas, $a_p$ is the number in $\zet_K/p^2$ that agrees with $z_i$ at each prime 
$\mathfrak{p}_i$ dividing $p$,
\begin{equation}
a_p = z_i \bmod \mathfrak{p}_i^2
\end{equation}
As in the previous examples, we have found this procedure (augmented with \eqref{lift} and 
\eqref{multiplicative} for $k$ not prime) such that via
\begin{equation}
\widehat{\calw}_A = \sum_d\tilde n_d q^d = \sum_d n_d \sum_k \frac{a_k}{k^2} q^{dk}
\end{equation}
it returns invariants $n_d$ with no $d^2$ in the denominator, up to some significant order
$d$.

To make the procedure easier to follow, we discuss as an example, the results for 
\begin{equation}
n_1 = 10000 \lambda (-7624+4517 \lambda^2)
\end{equation}
(For incidental reasons, we revert here to the non-integral generator $\lambda$. This does not
change the results.)
Clearly, under $\lambda\mapsto-\lambda$, $n_1\mapsto -n_1$, so $z=-1$. For any element $g\in G$ 
of the Galois group that reverses the sign under the square-roots in \eqref{square} (namely, 
rotation by $\pi/2$ (order 4) or horizontal or vertical flip (order 2)), the resulting $z= g(n_1)/n_1$
is a root of the polynomial
\begin{equation}
\eqlabel{galpo}
 583443+135146154523047386 z^2+583443 z^4
\end{equation}
Because $583443=3^5 \cdot 7^4$, but $7$ does not divide the discriminant of $K$, that prime will 
require a bit of a special treatment. 

For the first few non-trivial primes, we find the following table
\begin{equation}
\begin{array}{c|c|c|c}
p& P/5^3\bmod p & f_p:= (n_1)^{p-1}\bmod p & a_p \bmod p^2\\ \hline
7	& (2+\lambda) (5+\lambda) (1+\lambda^2)		& 2+3 \lambda^2 &40+20\lambda^2\\
11	& 8+4 \lambda^2+\lambda^4		& 2+7 \lambda^2 	&24+62\lambda^2\\
13	& 6+4 \lambda^2+\lambda^4		& 4+12 \lambda^2 	&43+25\lambda^2\\
17	& (7+\lambda)^2 (10+\lambda)^2		& {\it ramifies} 	& \\
19	& (7+\lambda^2) (16+\lambda^2)		& 18 			& -1 \\
23	& (1+\lambda) (8+\lambda) (15+\lambda) (22+\lambda)	& 1	& 1 \\
29	& 2+4 \lambda^2+\lambda^4		& 19+11 \lambda^2 	& 48+417\lambda^2\\
31	& (3+8 \lambda+\lambda^2) (3+23 \lambda+\lambda^2)	& 18+11 \lambda^2  &235+445\lambda^2\\
37	& (14+\lambda) (23+\lambda) (15+\lambda^2)&	3+20 \lambda^2 & 780+390\lambda^2 \\
\vdots & \vdots & \vdots  & \vdots
\end{array}
\end{equation}
How did we find the last column?
From the degree of the factors of $P\bmod p$, we may read off the order of the various
Frobenius elements. In most cases, this determines them completely: for inert primes 
such as $11$, $13$, $29$, the Frobenius has order $4$, so must be rotation by $\pi/2$ 
in \eqref{square}. For primes that split completely such as $p=23$, the Frobenius is 
trivial (as we were happy to learn some time ago!). For primes with one quadratic, and two 
linear factors, such as $7,37$, the Frobenius must be a diagonal flip. The only ambiguous cases are 
those with two quadratic factors, which could correspond to horizontal/vertical flip, or 
rotation by $\pi$.

It is easy to check that for $p=11,13,29,31$, $f_p$ solves \eqref{galpo} $\bmod p$,
and $a_p$ is simply the lift of that solution $\bmod p^2$. In particular, the Frobenius
at $31$ must be horizontal/vertical flip. 

For $p=19$, $f_{19}=-1$ at both factors, so Frobenius must be rotation by $\pi$. We
keep $a_{19}=-1\bmod 19^2$, just as we use $a_{23}=1$ since $23$ splits completely.

For $p=37$, $f_{37}=1$ at the linear factors, and $f_{37}=-1$ at the quadratic factor.
In other words, $f_{37}$ is a solution of $z^2-1=0\bmod 37$, and $a_{37}$ is the lift of 
that solution $\bmod 37^2$. 

What happened at $p=7$? From the structure of the factorization, it should be in the same
class as $p=37$. However, $(f_7)^2-1\neq 0\bmod 7$ \ldots\ldots\ Some reflection reveals 
that the denominator of $z$ in \eqref{galpo} being divisible by $7$ is due to the fact that 
$n_1$ vanishes at the two linear factors of $7$, so the action of the Frobenius automorphism 
as $(n_1)^{p-1}$ becomes completely ambiguous there. Independently however, we have known that 
the Frobenius should restrict to $1$ at the linear factors, and to $-1$ at the quadratic factor. 
This can be used to determine that $a_p = 5+6\lambda^2\bmod 7$, which as a solution of $z^2-1=0$ 
may then safely be lifted $\bmod 7^2$.

We may summarize the A-model discussion by stating that once again the Ooguri-Vafa multi-cover 
formula \eqref{oovast} has proven to be basically correct, but that it needs a significant 
refinement in arithmetically non-trivial situations, which we have encountered here for the first 
time. The refinement is provided by the D-logarithm, which we conjecture is an analytic function 
attached to individual algebraic numbers $n_d$. The sequences $(a_k)$ defining the D-logarithm are 
specified $\bmod k^2$ by studying the action of the Galois group on $n_d$. It remains to be seen
whether this remarkable structure can be sharpened and explained more fundamentally, and how
it ties in with the rest of our subject. In the remaining section \ref{discussion}, we will 
present some initial thoughts that make this not impossible.

\subsection{Main component}

The only purpose of this subsection is to point out that it is possible to 
calculate the large volume expansion of the superpotential also on the main
component of $\calh_{\rm conics}$, which we called $\cali_2$ in section
\ref{largecomplex}. Consider the third group in Table \ref{groups}.
We solve the Picard-Fuchs equation with inhomogeneity \eqref{solving}, apply
the mirror map, and obtain
\begin{equation}
\eqlabel{difficult}
\begin{split}
\frac{4\pi^2}{(-\lambda)^{1/2}}\calw_A = & 
\textstyle
- 175 \lambda^5 (-5+\lambda^7)q^{1/7}- \frac{125 (-94+17 \lambda^7)}{4}q^{2/7}
-\frac{25 \lambda^2 (-70585+31748 \lambda^7)}{63}q^{3/7} \\
&\textstyle -\frac{25\lambda^4 (-2394125+191028 \lambda^7)}{2^4\cdot 7^2} q^{4/7}
- \frac{31 \lambda^6 (-245997065+63500311 \lambda^7}{5\cdot 7^3} q^{5/7} \\
&\textstyle
-\frac{5 \lambda (-8907388019619+2655707519021 \lambda^7)}{2^2 \cdot 3^3\cdot 7^4} q^{6/7}
- \frac{5\lambda^3 (-3595649177+980861072 \lambda^7)}{7} q\\
&\textstyle
-\frac{ \lambda^5 (-1905271484195274460+512248788482392343 \lambda^7)}{2^7 \cdot 7^7}q^{8/7}
+\cdots \end{split}
\end{equation}
where $\lambda$ is the algebraic number with minimal polynomial
\begin{equation}
\lambda^{14}-5\lambda^7+5 = 0
\end{equation}
It is difficult to obtain convincing tests of our general formalism from \eqref{difficult}, 
but the first few orders in the expansion are encouraging: $p=2,5,7$ divide the discriminant,
so are allowed in the denominator. Subtraction of $\tilde n_1$ from $\tilde n_3$ works as 
expected, with $a_3^{(1)}=8 \lambda^4+2 
\lambda^{11}$. I have no explanation for the apparent anomaly at $d=6$ (which has $3^3$ in the 
denominator).

\section{Discussion}
\label{discussion}

In this work, we have studied families of algebraic cycles on the mirror quintic 
represented by curves of low degree. We have obtained a fairly complete picture of 
those conics that deform with the mirror quintic. We have seen how the Newton-Puiseux 
expansion around large complex structure limit splits the algebraic cycle into groups, 
each governed by an algebraic number field. We have 
then calculated the truncated normal function (up to an additive constant) by solving 
the inhomogeneous Picard-Fuchs equation. The irrationality does not disappear after 
application of the mirror map, confirming a long-standing expectation. To exhibit the 
underlying (algebraic) integrality\footnote{One should work with a notion of 
integrality that requires a non-negative valuation at all primes except those that
ramify.} of the expansion, we have introduced the D-logarithm as an arithmetic twist 
of the di-logarithm. This formalism generalizes all previously known cases and we might 
expect that it is complete. Indeed, we formulate the main computational result of this paper 
as follows:

\noindent
{\bf Conjecture:} {\it The A-model ($q$-)expansion of the truncated normal function 
associated with an algebraic cycle takes the form
\begin{equation}
\eqlabel{takestheform}
\widehat \calw_A(q) = \sum_d \tilde n_d q^{d/r}
\end{equation}
were $r\in \zet_{>0}$, and the $\tilde n_d$ live in an algebraic number 
field $K$, with $d^2 \tilde n_d$ singular at most at the discriminant of $K$.

In the expansion
\begin{equation}
\eqlabel{singular}
\widehat\calw_A(q) = \sum_{d} n_d \,{\rm Li}_2^{\rm D} (q^{d/r})
\end{equation}
the $n_d$ themselves are singular at most at the discriminant of $K$. Here, the D-logarithm
\begin{equation}
{\rm Li}_2^{\rm D}(x) = \sum_{k=1}^\infty \frac{a_k}{k^2} x^k
\end{equation}
is an analytic function that may be attached to any such algebraic number $n_d$. 
The coefficients $a_k$ are determined $\bmod k^2$ by studying the action of
the Galois group on $n_d$.}

(We did not state all assumptions explicitly, such as that we are on a family of Calabi-Yau
threefolds, expand around a large complex structure limit, {\it etc.}. The allowed 
singularities include denominators whose order at the discriminant grows (say linearly)
with $d$. It should also be clear that we expect the same 
to work for higher-dimensional moduli spaces. The statements defining the 
D-logarithm are \eqref{pprime}, \eqref{lift}, \eqref{multiplicative}.)

It is possible that several ingredients for a proof of the above statements are contained
in the work of Vologodsky, Schwarz and Kontsevich, see \cite{scvo}.

Turning to possible interpretations of the result, we recall that the truncated normal 
function gives the contribution to the space-time superpotential of a D-brane configuration 
that specifies the algebraic cycle, in the B-model. By mirror symmetry, there should be 
an A-model setup that calculates the expansion \eqref{takestheform} directly. In our 
examples, such a setup would contain a Lagrangian submanifold $L\subset X$ of the quintic 
threefold, and the $q$-expansion should be the expansion in worldsheet instantons of disc 
topology.

We lack the tools to exhibit such A-branes directly, but we can nevertheless try to understand 
whether there is room for the various ingredients: a number field $K$ governing the vacuum
structure of $L$, an action of the Galois group $G$, and instanton contributions that 
evaluate to algebraic numbers $\tilde n_d\in K$.

(i) The irrationality of the instanton contribution is insofar surprising as it has not
been seen in any previous example. If anything, $\tilde n_d$ should be open Gromov-Witten
invariants counting holomorphic maps $(D,\del D)\to (X,L)$. In all cases studied
so far, such invariants always evaluated to rational numbers. Mathematically, the 
counts are given by intersection theory on moduli spaces $\overline{\calm}$ of stable maps 
as integrals against the virtual fundamental class,
\begin{equation}
\eqlabel{letalone}
\tilde n_d \overset{?}{\sim} \#\{u:(D,\del D)\to (X,L)\} 
\overset{?}{\sim} 
\int_{[\overline{\calm}]^{\rm virt}} {\bf 1}
\end{equation}
In general however, open Gromov-Witten invariants have not actually been defined, let 
alone does there exist a formula like \eqref{letalone}. The two exceptions are toric
manifolds \cite{kali} and anti-holomorphic involutions \cite{jake}. The main obstacle 
to doing this in general 
has long been recognized to be the presence in $\overline{\calm}$ of boundaries in 
real co-dimension one. It is not clear therefore whether there actually exists a good
invariant intersection theory on these spaces.

(ii) From the world-sheet point of view, formulas such as \eqref{letalone} arise
as the reduction of the path-integral to the finite-dimensional space of zero-modes:
roughly speaking, because of the vanishing of the fermion kinetic term in the action, 
one has to pull down the four fermion interaction involving the curvature. In many, 
favorable, situations, the resulting bosonic integrals have a cohomological 
interpretation in terms of intersection theory. However, except perhaps with large 
amounts of supersymmetry, there is no {\it a priori} reason why this should 
happen. It is quite conceivable that in the presence of boundaries, we do not have a
strict intersection theoretic interpretation, but the integral still makes sense,
and calculates a kind of ``volume'' of the moduli space. Such a volume could very
well evaluate to an algebraic number.

(iii) For a related thought, we recall that ordinary Gromov-Witten invariants are
in general not integer, but rational, numbers because of the presence of certain kinds
of orbifold singularities in the moduli space. The denominators are the orders of
the corresponding identification groups. In the context of open Gromov-Witten theory,
the moduli spaces could have other kinds of singularities, such as boundaries
and corners, and in particular the latter could potentially make arbitrary irrational
contributions.

(iv) Of course, those two options, (ii) and (iii), assume that open Gromov-Witten 
theory exists in general, and defines actual invariants with an ``enumerative'' 
meaning. An alternative 
attitude is that any such definition will depend on arbitrary choices (a common
examples being the framing ambiguity of ref.\ \cite{akv,kali}). With a superpotential
interpretation for the open Gromov-Witten invariants, this would mirror the
issue, discussed in the introduction, that only the on-shell values of the space-time
superpotential have an invariant meaning independent for example of field 
redefinitions. In this interpretation, the invariants $\tilde n_d$ would be
irrational because they are {\it on-shell} and {\it invariant}, whereas the actual 
(rational) counts of discs would happen {\it off-shell}, and {\it not be invariant}.

(v) This way of looking at the situation is perhaps best suited for explaining
how the field extension could arise in the A-model. At the beginning, the underlying Lagrangian 
$L$ might have non-trivial topology and deformations, which get lifted by those very
worldsheet instantons that we are trying to count. Intuitively, the critical 
points of the superpotential correspond to points in the moduli space of $L$ at
which the worldsheet instantons are ``balanced'' against each other. If there
are sufficiently many discs of comparable area, then these critical point 
conditions will select some general irrational points in the classical moduli 
space of $L$. If finitely many discs are relevant for this problem, and off-shell
counts are rational, then we should be dealing with a finite algebraic extension 
of the rationals. (At the moment, I do not see how to get infinite, or transcendental
extensions in the B-model.)

(vi) Ideally, one would like to understand this in a suitable local model. For
customary toric Lagrangian branes however, there are at most two discs determining
the critical points, as for the conifold \eqref{conifold}. A local model
realizing a non-trivial field extension is therefore not likely to be toric (and
in a sense would not be fully local since disc instantons ending in different places
on $L$ would be relevant). 

(vii) A local model would also be desirable in order to understand the structure
of the multi-cover formula \eqref{singular}. Otherwise, we have comparably little 
to offer for interpreting the invariants $n_d$. From the previous discussion in 
ref.\ \cite{oova}, we expect a relation to the spectrum of appropriate BPS states 
(solitons) interpolating between the supersymmetric vacua. Given that the latter 
are in correspondence with roots of a polynomial equation, the Galois group of the 
relevant number field will act also on those BPS states. An irrational ``dimension''
could be part of the package of these Galois representations. The $a_k$ would then
be other traces, and the formula \eqref{singular} could perhaps be understood by 
revisiting the derivation in \cite{oova} in light of such results.

For a very brief sampling of other recent works on various ways to relate geometry 
and physics of Calabi-Yau threefolds with number theory, see 
\cite{gmoore,candelas,vanstraten,noriko,vangeemen,yang}

\begin{acknowledgments}
I am grateful to 
Henri Darmon,
Hans Jockers,
Sheldon Katz,
Matt Kerr,
Josh Lapan,
Wolfgang Lerche,
Greg Moore,
David Morrison,
and Noriko Yui
for valuable discussions, comments, and encouragement.
I thank Anca Musta\c t\v a for sharing the results of \cite{mustata}.
I would like to thank the KITP in Santa Barbara for sunny 
hospitality during the tedious writing of section \ref{conics}. Some
of the results presented here were also announced at the BIRS Workshop
on ``Number theory and physics at the crossroads'', May 8--13, 2011.
Special thanks to Henri Darmon for help in unraveling the D-logarithm.
This research is supported in part by an NSERC discovery grant and a
Tier II Canada Research Chair. This research was supported in part by 
DARPA under Grant No. HR0011-09-1-0015 and by the National Science 
Foundation under Grant No.\ PHY05-51164.
\end{acknowledgments}





\end{document}